\documentclass{aa}

\usepackage{graphicx}
\usepackage{txfonts}
\usepackage{natbib}
\usepackage{soul}
\usepackage[dvipsnames]{xcolor}
\usepackage{tablefootnote}

\graphicspath{{./figure/}}

\begin{document} 

\title{Effects of the chromospheric Ly$\alpha$ line profile shape on the determination
   of the solar wind \ion{H}{I} outflow velocity using the Doppler dimming technique}
\author{G.~E. Capuano\inst{1}
        \and
        S. Dolei\inst{2}
        \and
        D. Spadaro\inst{2}
        \and
        S.~L. Guglielmino\inst{1,2}
        \and
        P. Romano\inst{2}
        \and    
        R. Ventura\inst{2}
        \and
        V. Andretta\inst{4}
        \and
        A. Bemporad\inst{3}
        \and
        C. Sasso\inst{4}
        \and
        R. Susino\inst{3}
        \and
        V. Da Deppo\inst{5}
        \and
        F. Frassetto\inst{5}
        \and
        S.~M. Giordano\inst{3}
        \and
        F. Landini\inst{3}
        \and
        G. Nicolini\inst{3}
        \and
        M. Pancrazzi\inst{3}
        \and
        M. Romoli\inst{6}
        \and
        L. Zangrilli\inst{3}
        }

        \institute{Dipartimento di Fisica e Astronomia ``Ettore Majorana'' -- Sezione Astrofisica, Universit\`a degli Studi di Catania, Via S.~Sofia 78, I-95123 Catania, Italy \\
        \email{giuseppe.capuano@inaf.it}
        \and
        INAF - Osservatorio Astrofisico di Catania,
        Via S.~Sofia 78, I-95123 Catania, Italy
        \and
        INAF - Osservatorio Astrofisico di Torino,
        Via Osservatorio 20, I-10025 Pino Torinese (TO), Italy
        \and
        INAF - Osservatorio Astrofisico di Capodimonte,
        Salita Moiariello 16, I-80131 Napoli, Italy
        \and
        CNR - Istituto di Fotonica e Nanotecnologie,
        Via Trasea 7, I-35131 Padova, Italy
        \and
        Dipartimento di Fisica e Astronomia, Universit\`a degli Studi di Firenze,
        Largo Enrico Fermi 2, I-50125 Firenze, Italy
        }
                
   \date{}

 \abstract 
  {The determination of solar wind \ion{H}{I} outflow velocity is fundamental to shedding
  light on the mechanisms of wind acceleration occurring in the corona.
  Moreover, it has implications in various astrophysical contexts, such as
  in the heliosphere and in cometary and planetary atmospheres.}
  {We aim to study the effects of the chromospheric
  Ly$\alpha$ line profile shape on the determination of the outflow speed of coronal
  \ion{H}{I} atoms via the Doppler dimming
  technique.
  This is of particular interest in view of the
  upcoming measurements of the Metis coronagraph aboard the Solar
  Orbiter mission.}
  {The Doppler dimming technique exploits the decrease of coronal Ly$\alpha$
  radiation in regions where \ion{H}{I} atoms flow out in the solar wind. Starting from
  UV observations of the coronal Ly$\alpha$ line from the UltraViolet Coronagraph Spectrometer (UVCS), aboard the Solar and Heliospheric Observatory (SOHO), and simultaneous measurements
  of coronal electron densities from {\it{pB}} coronagraphic observations, we explored the effect
  of the profile of the pumping chromospheric Ly$\alpha$ line. We used measurements from the Solar UV Measurement of Emitted Radiation (SUMER), aboard SOHO, the Ultraviolet Spectrometer and Polarimeter (UVSP), aboard the Solar Maximum Mission (SMM), and the Laboratoire de Physique Stellaire et Planetaire (LPSP), aboard the Eight Orbiting Solar Observatory (OSO-8), both from representative
  on-disc regions, such as coronal holes and quiet Sun and active regions, and as a
  function of time during the solar activity cycle.
  In particular, we considered the effect of four chromospheric line parameters: line
  width, reversal depth, asymmetry, and distance of the peaks.}
  {We find that the range of variability of the four line parameters
  is of about 50\% for the width, 69\% for the reversal depth,
  and 35\% and 50\% for the asymmetry and distance of the peaks, respectively.
  We then find that the variability of the pumping Ly$\alpha$ profile
  affects the estimates of the coronal \ion{H}{I} velocity by about 9-12\%. This uncertainty
  is smaller than the uncertainties due to variations of other physical quantities,
  such as electron density, electron temperature, \ion{H}{I} temperature, and integrated
  chromospheric Ly$\alpha$ radiance.}
  {Our work suggests that the observed variations in the chromospheric Ly$\alpha$ line profile parameters
  along a cycle and in specific regions negligibly affect the determination of the solar wind speed of
  \ion{H}{I} atoms. Due to this weak dependence, a unique shape of the Ly$\alpha$ profile over the solar disc that is constant in time
  can be adopted to obtain the values of the solar wind \ion{H}{I} outflow velocity.
  Moreover, the use of an empirical analytical chromospheric profile of the Ly$\alpha$,
  assumed uniform over the solar disc and constant in time, is justifiable in
  order to obtain a good estimate of the coronal wind \ion{H}{I} outflow
  velocity using coronagraphic UV images.}


   \titlerunning{Effects of the Ly$\alpha$ profile on the \ion{H}{I} outflow velocity determination}
   
   \maketitle
%

\section{Introduction}
The solar wind is a tenuous plasma that continuously escapes from the Sun
into the interplanetary medium.
It extends for thousands of millions of kilometers and is mainly characterised by a fast
component, with typical velocities of about $500 - 800  \,\mathrm{km \,s}^{-1}$ flowing
from coronal holes (CHs), and a slow component, with typical velocities of
about $300 - 500  \,\mathrm{km \,s}^{-1}$, mainly related to the streamer belt regions
\citep[see e.g.][]{McComas1998,McComas2008}.

The high coronal temperatures, of which the origin is still unclear,
cause wind expansion \citep[e.g.][]{Parker1958}
and eventually the acceleration of slow and fast solar wind.
It is thought that such acceleration could be the result of, for example,
energy dissipation through the ion cyclotron resonance of high-frequency left-hand polarised
Alfvén waves \citep[see e.g.][]{Ofman2010}.

Given the high coronal temperatures, the solar wind is mainly composed
of fully ionised hydrogen and helium and of free electrons.  The latter
component is mainly observed through the scattering and polarisation
of the white light coming from the photosphere (Thomson scattering).
Other ions of heavier minor elements are present in smaller amounts.

The ESA/NASA Solar and Heliospheric Observatory \citep[SOHO;][]{Domingo1995} spacecraft
returned a wealth of data, which provided a new understanding of the physical phenomena
that contribute to the acceleration of the solar wind in the corona. In particular, the
UltraViolet Coronagraph Spectrometer \citep[UVCS;][]{Kohl1995} on board SOHO allowed
us to obtain crucial results, such as information on elemental abundances and
kinetic temperatures of \ion{H}{I} and \ion{O}{VI} ions in different coronal
structures. \citet{Antonucci2005}, \citet{Susino2008}, and \citet{Abbo2010a} suggested
that the main source of the slow solar wind is the boundary region between CHs
and streamers, where the wind is dependent on the magnetic field topology.
Outflows from active regions (ARs) could even contribute to the slow wind
component \citep{Zangrillipoletto2016}.
The presence of pseudo or unipolar streamers can also contribute to the slow
solar wind. Furthermore, white light observations with the Large Angle and
Spectrometric COronagraph \citep[LASCO;][]{Brueckner1995} aboard SOHO
have shown that outflows are connected to both large-scale 'streamer blowout'
structures, the coronal mass ejections (CMEs) closely associated
with magnetic reconnection at the current sheet above the cusp of the streamers,
and to small-scale inhomogeneities (blobs), linked to quasi-periodic
emission of plasma from cusps of helmet streamers
(\citealp{Sheeley1997,Wang1998,Song2009,ViallVourlidas2015};
see also the review of \citealp{Abbo2016}).

Generally, observations of emission lines in the off-limb corona, such as the
\ion{H}{I} Ly$\alpha$ (121.6~nm), which are mainly due to scattering processes of
radiation coming from the low solar atmosphere, can be used to determine the radial component of the outflow
velocity ($v_w$) of the specific atom or ion that generates the scattering
through the Doppler dimming technique \citep{HyderLites1970,Withbroe1982,Noci1987}. This technique consists of the
analysis of the intensity decrease of the coronal ultraviolet (UV) radiation
coming from flowing regions, and it relies on some assumptions and prior knowledge
of the geometrical and physical properties that are representative of the
coronal environment (such as electron density, electron temperature, and kinetic
temperature of the scattering ions) and of the source of the UV radiation
scattered (such as intensity and line profile of the exciting chromospheric radiation).
However, in those cases in which observations come from coronagraphs designed
for imaging only, these quantities cannot be directly determined and they must
 be taken from the literature. Critical works aimed at investigating
how the choice of the parameters adopted from the literature (and their uncertainties)
or from specifically constructed databases may affect the results of the
Doppler-dimming analysis have been presented very recently.

In this context, \citet{Dolei2016} analysed the \ion{H}{I} Ly$\alpha$ line
profiles observed by UVCS instrument in a number of polar, mid-latitude, and
equatorial structures. They considered a large amount of data acquired in a
time range longer than a complete solar cycle (1996 -- 2012) and determined
the temperature components for neutral hydrogen at different phases of solar
activity. They were able to derive the \ion{H}{I} temperature radial profiles
for several heliocentric distances (from $1.3 \,R_{\odot}$ to $4.5 \,R_{\odot}$)
and polar angles.

\citet{Dolei2018} used coronagraphic Ly$\alpha$ observations from UVCS and
visible light (VL) images obtained with LASCO and the Mark-III Coronameter
\citep[Mk3/MLSO;][]{Fisher1981} to study the dependence of the derived
solar wind \ion{H}{I} outflow velocity on the physical parameters that
characterise the scattered coronal Ly$\alpha$ emission when the Doppler
dimming effect is considered.
They found that the electron temperature has the most important impact on
the determination of the \ion{H}{I} outflow velocity, while the coronal
\ion{H}{I} temperature has an important role at larger heliocentric distances,
where higher speeds are reached.

\citet{Dolei2019} also used UVCS and LASCO-Mk3 observations in order to
analyse the effect of chromospheric radiation inhomogeneities on the solar
wind \ion{H}{I} outflow velocity determination with respect to the case
in which a uniform-disc brightness approximation is adopted. In particular,
they created a Carrington map of the non-uniform solar chromospheric Ly$\alpha$
radiation using a correlation function between the \ion{H}{I} 121.6 nm and
\ion{He}{II} 30.4 nm intensities \citep{Auchere2005}, while some approximations
concerning the coronal \ion{H}{I} temperature and a constant value of the
radiance over the entire solar disc were adopted for the uniform-disc brightness case.
They found that the uniform condition leads to an overestimated velocity of
$50-60  \,\mathrm{km \,s}^{-1}$ in polar and mid-latitude regions, while
underestimated velocity values are obtained at the equatorial regions.
This difference decreases at higher altitudes.

The present work can be considered as a step forward in the investigations
carried out by \citet{Dolei2018,Dolei2019}.
The aim of this paper is to study the effects induced on the determination of
the solar wind \ion{H}{I} velocity by the variation of the exciting chromospheric
Ly$\alpha$ line profile shape, which can occur among different regions on the solar
surface, such as quiet or active regions and CHs, or during different phases
of the solar activity cycle. For this purpose, we considered Ly$\alpha$
chromospheric observations performed by the Solar UV Measurement of Emitted Radiation
\citep[SUMER;][]{Wilhelm1995} aboard the SOHO spacecraft, which include full disc
observations \citep[][]{Lemaire2015} and the best sets of Ly$\alpha$ solar-disc
observations currently available \citep{Curdt2008,Tian2009a,Tian2009b}. We also
considered other observations concerning ARs, quiet Sun (QS) regions, and an equatorial
CH performed by the Ultraviolet Spectrometer and Polarimeter \citep[UVSP;][]{Miller1981}
aboard the Solar Maximum Mission \citep[SMM;][]{Woodgate1980} and the LPSP instrument
\citep[Laboratoire de Physique Stellaire et Planetaire;][]{Artzner1977,Bonnet1978}
aboard the Eight Orbiting Solar Observatory (OSO-8).
Other parameters critically affecting the coronal Ly$\alpha$ emission,
such as the electron temperature, the electron density, and the neutral
hydrogen temperature, are the same as those adopted by \cite{Dolei2019}. 

In order to facilitate the comparison, bidimensional (2D) maps of outflow solar
wind \ion{H}{I} velocity were produced by using the same numerical code as in
\citet{Dolei2018,Dolei2019}, based on the reproduction of the synthetic coronal
Ly$\alpha$ intensity along the line of sight (LOS). The numerical computations have
allowed us to change the value of the wind velocity $v_w$ until a match with the
observed coronal intensity is reached. Moreover, this allowed us to estimate
the solar wind velocity as a function of heliocentric distance and latitude.
In obtaining each map of solar wind \ion{H}{I} outflow velocity, we assumed
that the chromospheric intensity and the line profile of the Ly$\alpha$ emission
are unique over the entire solar disc.

Our study is of particular interest for the analysis and interpretation of coronal
Ly$\alpha$ observations that will be obtained by the Metis coronagraph
\citep{Antonucci2020,Fineschi2020} onboard the Solar Orbiter spacecraft
\citep{Muller2013,Muller2020}.
Indeed, Metis, which is designed for imaging only, observes
linearly polarised VL (580 nm - 640 nm) and UV Ly$\alpha$ emission at the same time with a
field of view (FOV) that ranges from $1.6 \,R_{\odot}$ to $7.5 \,R_{\odot}$ \citep{Antonucci2020},
viewing the Sun from latitudes up to about $30^{\circ}$ with respect to the
equatorial plane \citep{Zouganelis2020}.

In Sect.~\ref{sec:dataanalysis}, the main mechanisms concerning the coronal emission,
the analysis technique used, and the observations are described.
Sect.~\ref{sec:res} is dedicated to the obtained results. Finally, in
Sect.~\ref{sec:discus}, we discuss the results and give our conclusions.

\section{Data analysis} 
\label{sec:dataanalysis}

\subsection{Mechanisms of coronal emission and the Doppler dimming phenomenon}
\label{subs:mec}
In a static corona, the emission coming from coronal atoms/ions is due to the
excitation induced by two main mechanisms: the collision with free electrons
and the resonant scattering of chromospheric radiation.
However, in the case of the \ion{H}{I} Ly$\alpha$ emission,
the collisional term contributes only to a
small fraction of the total coronal emission \citep[][Landi Degli'Innocenti;
private communication, quoted in \citealp{Noci1987}]{Gabriel1971,Raymond1997},
given the stronger dependence of the collisional component $\mathcal{I}_{col}$
of the coronal intensity on the electron column density
$N_e$ (i.e. $\mathcal{I}_{col}\propto{N_e^2}$) with respect to the resonant
one $\mathcal{I}_{rad}$ \citep[i.e. $\mathcal{I}_{rad}\propto{N_e}$; see e.g.][]{Withbroe1982}
and the low-density conditions present in the outer corona.
For this reason, it can be neglected.

Under the hypothesis of low-density coronal plasma, the radiance
$\mathcal{I}_{rad}$ (in units of erg cm$^{-2}$ s$^{-1}$ sr$^{-1}$)
coming from \ion{H}{I} atoms in a coronal point $P$ and measured
along the ${\bf{n}}$ direction, corresponding to the line of sight,
can be expressed by the following:

\begin{multline}
        \mathcal{I}_{rad}=\frac{0.833~h~B_{12}}{4\pi\lambda_{0}}\int_{-\infty}^{+\infty}{n_{e}~R_{\ion{H}{I}}(T_e)}~dl \\
        \times{\, \int_{\Omega}\frac{11+3({\bf{n}}\cdot{\bf{n'}})^2}{12}}~{F({\bf{n'}},v_w, \theta)}~d\Omega,
        \label{eq:Irad}
\end{multline}

\noindent
where $h$ is the Planck constant; $B_{12}$ is the Einstein
coefficient for the Ly$\alpha$ transition; $\lambda_{0}=121.567$~nm
is the central wavelength of the considered transition; $n_e$ is the
electron density; $0.833$ is the ratio between proton and electron
density for a gas that is completely ionised and contains 10$\%$ helium;
$T_e$ is the electron temperature; $R_{\ion{H}{I}}(T_e)$ is the total
hydrogen ionisation fraction as a function of $T_e$, meaning
the ratio between the neutral hydrogen and proton density \citep[see e.g.][]{Withbroe1982};
$dl$ is the infinitesimal path along the line of sight; ${\bf{n}'}$ is
the direction along which the chromospheric radiation reaches the coronal
point $P$; $\Omega$ is the solid angle under which the point $P$ subtends the solar disc
\citep[see e.g. Fig.~1 in][]{Dolei2015};$p(\Omega)=[11+3({\bf{n}}\cdot{\bf{n'}})^2] / [12~(4\pi)]$
is a geometrical factor that gives the angular dependence of the scattering
process \citep[see][]{BeckersChipman1974}; and

\begin{equation}
        F({\bf{n'}},v_w, \theta)=\int_{-\infty}^{+\infty}{I({\lambda'-\lambda_0-
        \delta\lambda}, {\bf{n'}})~\Phi({\lambda'-\lambda_0})~d{\lambda'}}
        \label{eq:equaz_doppler_factor}
\end{equation}

\noindent
is the convolution between the chromospheric specific intensity
$I(\lambda'-\lambda_{0}-\delta\lambda, {\bf{n'}})$ and the normalised
coronal absorption profile $\Phi({\lambda'-\lambda_0})$,
where $\lambda'$ is the wavelength.
The term

\begin{equation}
\delta\lambda=\frac{\lambda_{0}}{c}{\bf{v}}\cdot{\bf{n'}}=
\frac{\lambda_{0}}{c}v_w\cos{\theta}
\end{equation}

\noindent
is the Doppler shift seen along ${\bf{n'}}$ by the scattering atoms
due to the velocity ${\bf{v}}$ (supposed radially symmetric) of the
scattering atoms, where $v_w$ is the module of ${\bf{v}}$, $\theta$
is the angle between ${\bf{v}}$ and ${\bf{n'}}$, and {\it c} is the
speed of the electromagnetic radiation in the vacuum
(see Fig.~1 in \citealp{Dolei2015} for a clear description of the
geometry of the scattering process).
The specific intensity $I(\lambda'-\lambda_{0}-\delta{\lambda}, {\bf{n'}})=
I({\bf{n'}})\cdot\Psi(\lambda'-\lambda_{0}-\delta{\lambda})$ is the
product between the radiance $I({\bf{n'}})$ (in units of erg cm$^{-2}$ s$^{-1}$ sr$^{-1}$)
and the normalised intensity profile $\Psi(\lambda'-\lambda_{0}-\delta{\lambda})$
(in units of \AA$^{-1}$) of the chromospheric Ly$\alpha$ line.
Then, $F({\bf{n'}},v_w, \theta)$ gives information about the dependence
on the \ion{H}{I} atom's velocity of the radiative excitation efficiency
by chromosperic radiation.
If we assume that the velocity distribution of the absorbing \ion{H}{I}
coronal atoms is Maxwellian, the normalised coronal absorption profile
$\Phi(\lambda'-\lambda_{0})$ has a width that depends on the thermal motion
only (neglecting all contributions coming from non-thermal phenomena, such
as turbulence, waves, oscillations, and so on), given by

\begin{equation}
\Delta\lambda_D=\frac{\lambda_{0}}{c}\sqrt{\frac{2~k_B~T_{\ion{H}{I}}}{m_H}}
,\end{equation}

\noindent
where $k_B$ is the Boltzmann constant, $T_{\ion{H}{I}}$ is the \ion{H}{I}
temperature, and $m_H$ is the mass of hydrogen atoms.

\begin{figure}[t]
        \centering
        \includegraphics[width=\hsize]{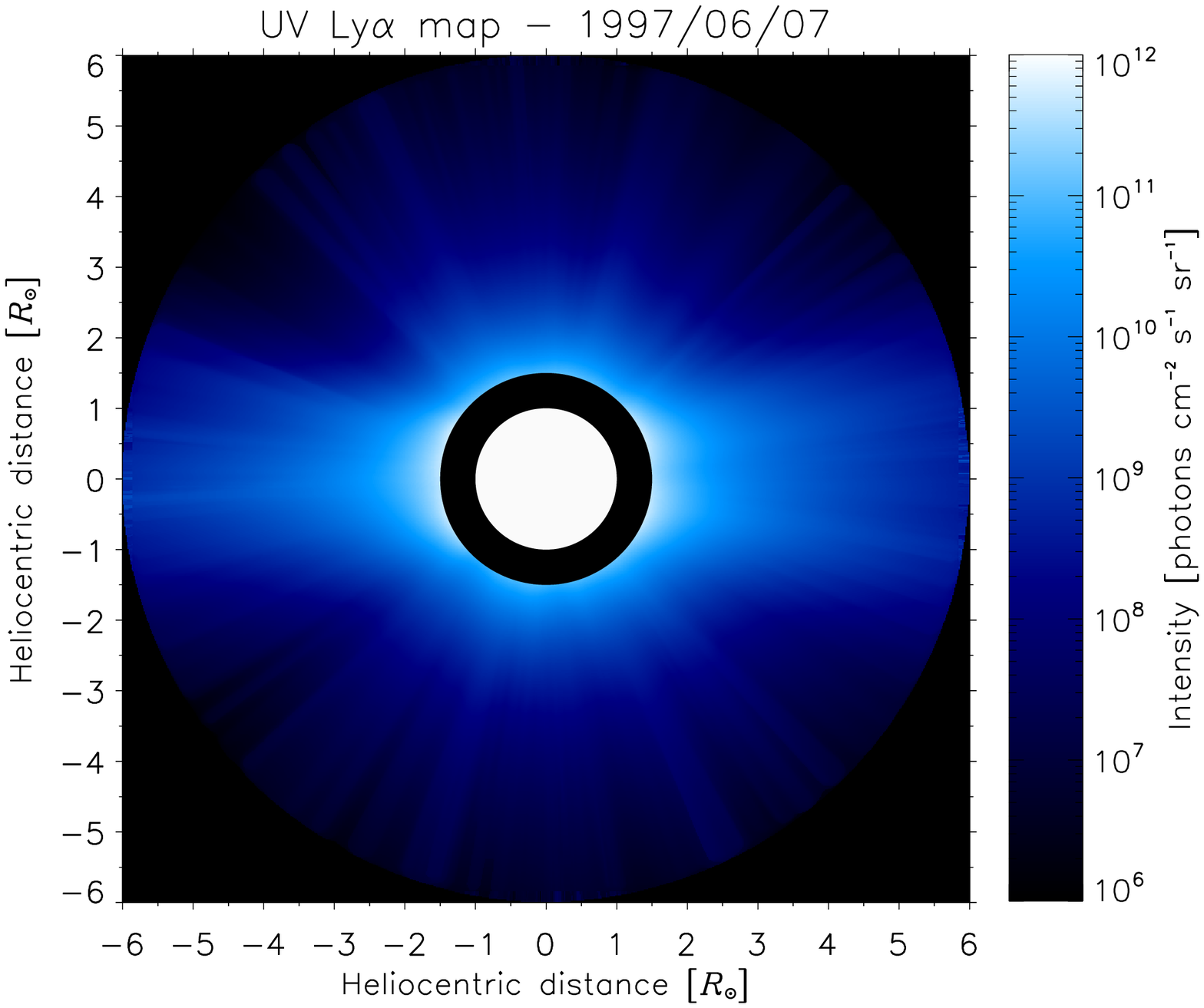}
        \includegraphics[width=\hsize]{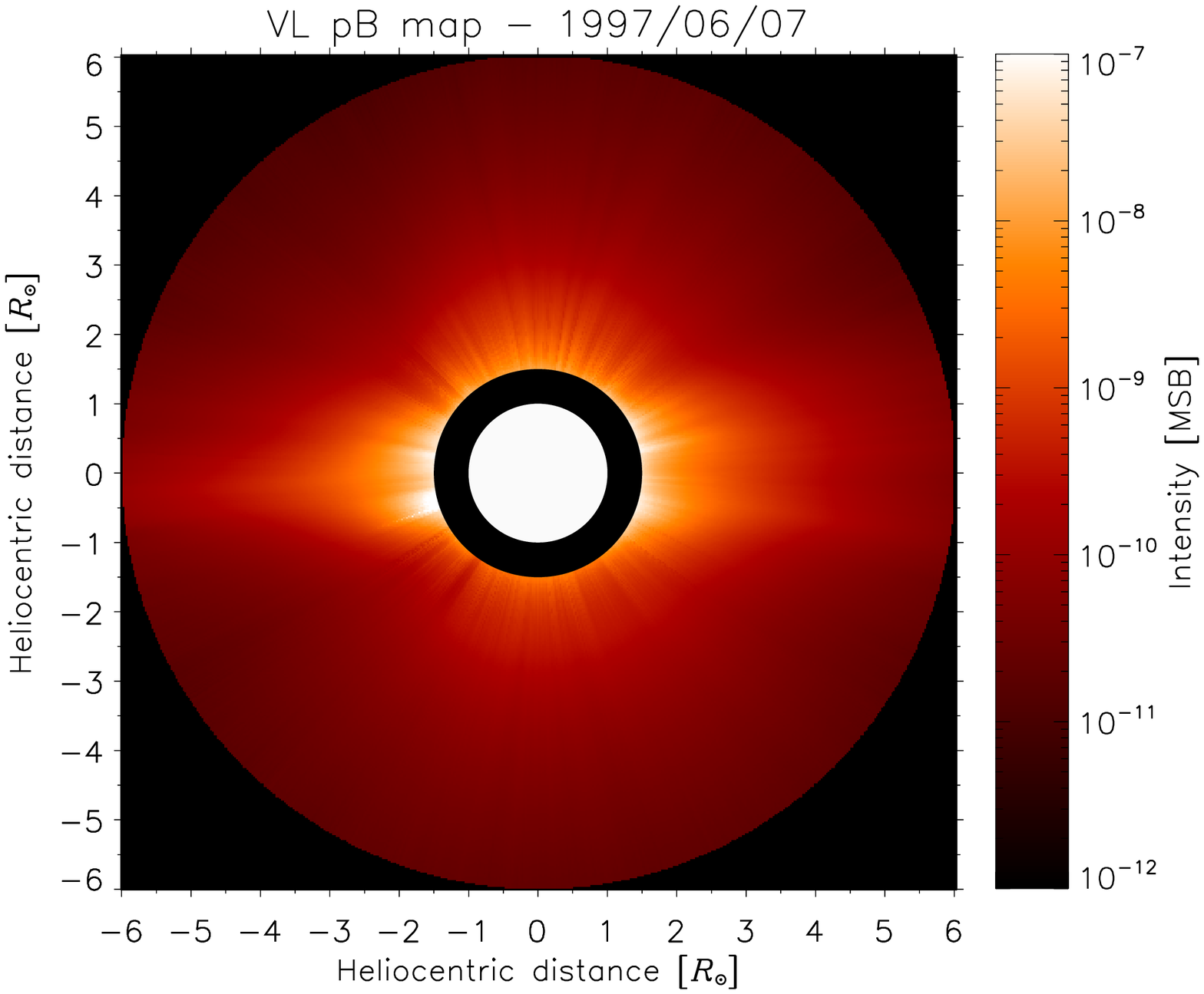}
        \caption{
        Bidimensional maps of UV coronal Ly$\alpha$ intensity (top panel),
        acquired by UVCS/SOHO, and VL {\it pB} emission (bottom panel), obtained
        from combined observations by the LASCO/SOHO and Mk3/MLSO instruments
        as a function of heliocentric distance and latitude. The maps have a
        FOV between $1.5 \,R_{\odot}$ and $6.0 \,R_{\odot}$, and refer to
        observations taken on June 7, 1997.
        }
        \label{fig:uv_pb}
\end{figure}

We also define the 'Doppler factor' $D(v_w)$ \citep[see][]{Noci1987}:

\begin{equation}
D(v_w)=\frac{\int_{\Omega}~F({\bf{n'}},v_w, \theta){p(\Omega)}~d\Omega}{\int_{\Omega}~F({\bf{n'}},v_w=0, \theta){p(\Omega)}~d\Omega}
\label{eq:ddf}
,\end{equation}

\noindent
which accounts for the overlapping between the coronal absorption profile and the
exciting chromospheric one as a function of $v_w$.

When the radiative resonantly scattered emission dominates, the outwards expansion
of the corona generates a decrease (dimming) in the Ly$\alpha$ coronal emission.
This phenomenon can be explained considering the Doppler redshift which an observer
positioned in $P$ measures in the chromospheric radiation. In fact, when the corona
is in a static condition, the profile of the chromospheric Ly$\alpha$ is centred at
$\lambda_{0}$, causing the maximum intensity of the scattered Ly$\alpha$ radiation
($D = 1$). However, when the corona is expanding, the scattered Ly$\alpha$ is dimmed
because of the Doppler shift of the chromospheric photons seen by the outflowing coronal
\ion{H}{I} atoms. The intensity of the scattered radiation tends to vanish when the
outflow velocity becomes higher, because the chromospheric and coronal profiles no
longer overlap ($D \approx 0$). Thus, it is possible to match the synthetic intensity
$\mathcal{I}_{rad}$ and that measured by observations by numerically tuning the $v_w$ value,
estimating in turn the outflow velocity.

\subsection{Observations}
\label{subs:obs}

\begin{figure}[t]
   \centering
   \includegraphics[width=\hsize]{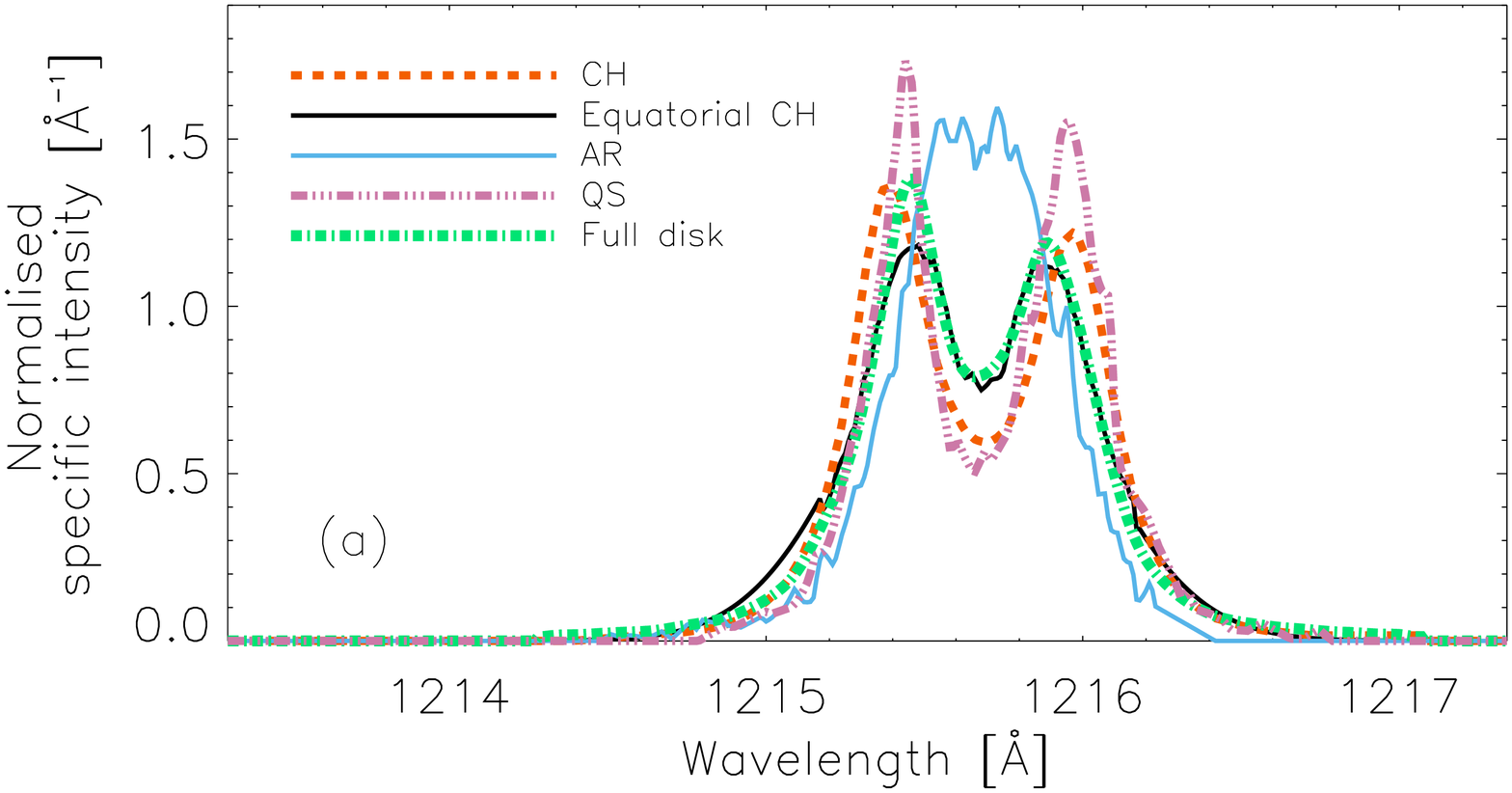}
   \includegraphics[width=\hsize]{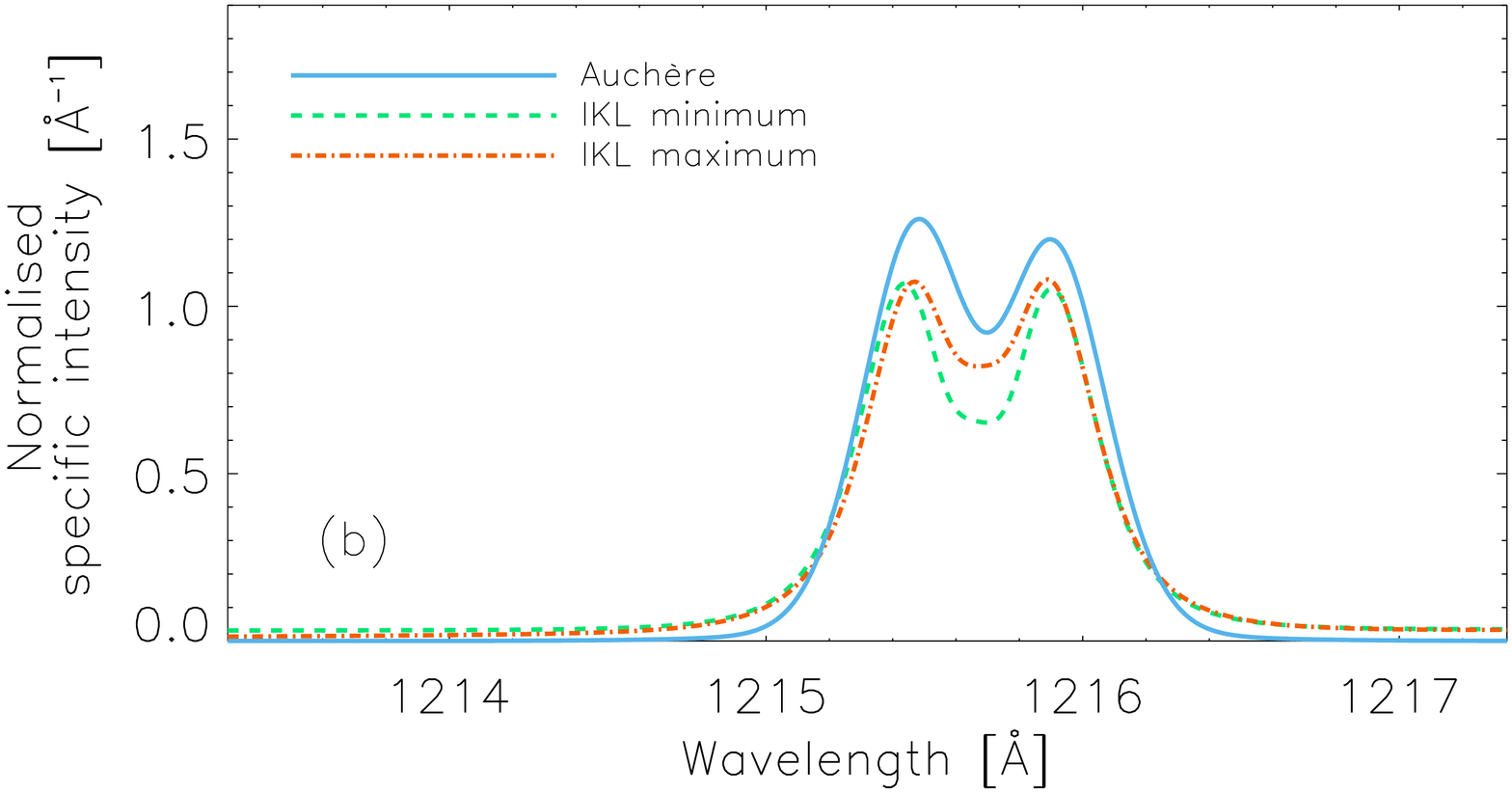}
      \caption{
      Panel ({\it a}): Ly$\alpha$ irradiance profiles reported in
      Fig.~6a and Fig.~12 of \citet{Fontenla1988} (QS and AR, respectively),
      as well as profiles observed on April 17, 2009 \citep[CH,][]{Tian2009b},
      on October 28, 1996 \citep[full disc,][]{Lemaire2015} and in an equatorial
      CH at the central meridian between November 27 and 29, 1975 \citep{BocchialiniVial1996}.
      Panel ({\it b}): Parametrised Ly$\alpha$ profiles from \citet{ikl2018}
      (IKL; at maximum and at minimum) and \citet{Auchere2005}.
      Each profile has been normalised to its total intensity.
      }
      \label{fig:profili_osservati_param}
\end{figure}

We describe the observations adopted in our analysis here.

\subsubsection{UV and {\it pB} observations}
A synoptic UV map related to the observations of coronal Ly$\alpha$ intensity
and analysing data acquired on June 7, 1997
(during the minimum of solar activity) by UVCS was obtained.
This 2D map was constructed, as explained by \citet{Bemporad2017} and
\citet{Dolei2018}, starting from synoptic UVCS observations and by performing a
power-law interpolation of intensities observed at different latitudes and altitudes.
A polarised brightness {\it pB} map in visible light was obtained by interpolating
observations acquired with Mk3 between $1.1 \,R_{\odot}$ and $1.5 \,R_{\odot}$ and
by LASCO above $2.5 \,R_{\odot}$ on the same day \citep[see][for more details]{Bemporad2017,Dolei2018}.
The maps, which are shown in Fig.~\ref{fig:uv_pb}, are those considered in the work by \citet{Dolei2019}.

It is worth mentioning that we only considered coronal observations related to the
minimum solar activity, although the analysed chromospheric observations, as
described below, concern an entire solar cycle.
However, since the aim of this work is to investigate the effects induced by the
chromospheric Ly$\alpha$ profile shape variation on the \ion{H}{I} outflow velocity,
it is not unreasonable to adopt a simplified scenario.

\subsubsection{Chromospheric observations of Ly$\alpha$ line profiles}
In this study, we aim to explore the impact of possible variations in the spectral
profile of the pumping chromospheric radiation, such as those caused by the solar
activity cycle, on the determination of $v_w$. For this purpose, we
considered full disc, QS and CH observations of the chromospheric Ly$\alpha$ line profiles
carried out by SUMER/SOHO during the solar activity cycle 23, ARs and QS
profiles acquired by UVSP/SMM, and profiles referring to an equatorial CH and a flare
acquired by the LPSP instrument aboard OSO-8, in order to take into account
all the possible chromospheric regions emitting the exciting radiation.

First, we used measurements of the Ly$\alpha$ line profile
relative to the entire solar disc. These were provided by \citet{Lemaire2015} and
are available at the Centre de Données astronomiques de Strasbourg
(CDS)\footnote{\url{https://cdsarc.unistra.fr/viz-bin/cat/J/A+A/581/A26}; \\ 
see Fig.~5 and Table A.1 of \citet{Lemaire2015} for more details.}. These
data, comprising 43 profiles, do not provide information about possible
variations of the chromospheric Ly$\alpha$ profile between different
regions on the solar surface observed at the same time. Nevertheless,
since they cover an entire solar cycle, it is reasonable to expect that
a good indication of the effect of the solar activity level on the shape
of the chromospheric Ly$\alpha$ line can be achieved, at least in the
coronal equatorial regions.

\begin{table*}
\caption{Four couples of selected observed profiles representing the extreme cases of the considered profile parameters.}
\label{tab:couples}
\centering
\begin{tabular}{c c | c c}
\hline\hline
\multicolumn{2}{c}{Line width}&\multicolumn{2}{c}{Reversal depth} \\ \hline
Narrowest&Broadest&Shallowest&Deepest\\
(AR 2340)&April 17, 2009 (CH)&(AR 2340)$^a$&(QS, Fig.~6a in \citealp{Fontenla1988})\\
0.58\AA&0.87\AA& -- &69\%\\
\hline
\multicolumn{2}{c}{Asymmetry of the peaks}&\multicolumn{2}{c}{Separation of the peaks} \\ \hline
$I\tiny{peak-blue}/I\tiny{peak-red}=1.16$&$I\tiny{peak-blue}/I\tiny{peak-red}=0.86$&Smallest separation&Largest separation \\ 
&&(Equatorial CH)&April 17, 2009 (CH) \\
October 28, 1996 (full disc)&(see text)&0.38\AA&0.57\AA \\
\hline
\begin{tabular}{l c c c}
\footnotesize{$a$) No reversal}
\end{tabular}
\end{tabular}
\end{table*}

Ly$\alpha$ profiles observed by SUMER in different areas of the solar disc
\citep{Curdt2008,Tian2009a,Tian2009b} represent the best Ly$\alpha$ solar-disc
observations currently available. We considered two profiles observed on
September 23, 2008 and on April 17, 2009, which are relevant to a QS region at the disc
centre and to a CH observed at the south pole, respectively
\citep[for more details, see][]{Tian2009a,Tian2009b}.

The UVSP profiles considered in this work were reported by \citet{Fontenla1988}
in their Figs.~3, 5, 6, and~12, where the geocoronal absorption was removed,
while the LPSP instrument on OSO-8 profiles were acquired in an equatorial
CH at the central meridian between 27 and 29 November, 1975 \citep{BocchialiniVial1996}
and in a faint flare that occurred in an AR observed on April 15, 1978 \citep{Lemaire1984}.
Analytic fits to chromospheric Ly$\alpha$ profiles were also taken into account.
In order to examine the variation of each parameter of the Ly$\alpha$ line profile
(line width, reversal depth, asymmetry of the peaks, and separation of the peaks),
we selected four couples of profiles, each of them representing extreme cases of
the considered profile parameters. In Table~\ref{tab:couples}, we report these
four couples, determined as described in the following. However, we ensured
that all the extreme cases analysed were represented by using only the five observed
profiles shown in panel ({\it a}) of Fig.~\ref{fig:profili_osservati_param}, where
each of them has been normalised to its own total intensity.

We measured the full width at half maximum (FWHM) of each selected Ly$\alpha$
chromospheric profile once these were normalised to their total intensities.
The mean value of the two peaks' highest intensities was adopted as the
maximum intensity of the Ly$\alpha$ line profile. Its half value was then used
to calculate the FWHM of the line. Only in the case of the considered profile
without reversal, meaning that relative to AR 2340 and reported in
Fig. 12 of \citet{Fontenla1988}, did we obtain the FWHM by performing a Gaussian fit.

We find a difference between the line widths of about 50\% at most over the entire
cycle, with respect to the narrowest observed Ly$\alpha$ chromospheric profile. 
It is worth noting that we did not find evidence of a correlation of the variation
of the line widths with the solar cycle. Indeed, the two line profiles that exhibit
the largest difference between their FWHMs correspond to that observed on
April 17, 2009 (CH) and that reported in Fig.~12 of \citet{Fontenla1988} (AR 2340), respectively.
These profiles are shown in panel ({\it a}) of Fig.~\ref{fig:profili_osservati_param} and in Fig.~\ref{fig:DF}
(red dashed line and blue solid line, respectively). They are considered extreme
cases with regard to the line profile width.

The shallowest and the deepest profiles in terms of reversal depth are those
described in Fig.~12 (AR 2340, showing a negligible reversal) and Fig.~6a (QS)
of \citet{Fontenla1988}, respectively. The depth was calculated by considering
the minimum central intensity value of the profile and dividing this by the mean
value between the maximum peak intensities. The reversal depth variation of the
deepest profile is equal to about 69\% with respect to the shallowest one.
These profiles, representing the extreme cases related to the profile reversal
depth, are displayed in panel ({\it a}) of Fig.~\ref{fig:profili_osservati_param}
(blue solid line and magenta dash-dot-dot-dot line, respectively).

The peak-to-peak asymmetry between the peaks of each profile was measured by
considering the ratio $I\tiny{peak-blue}/I\tiny{peak-red}$. The profile with
the largest difference between the maximum peak intensities is that observed on
October 28, 1996 (full disc), which exhibits the highest blue asymmetry
($I\tiny{peak-blue}/I\tiny{peak-red}=1.16$). This profile is displayed in
panel ({\it a}) of Fig.~\ref{fig:profili_osservati_param} and in panel ({\it c}) of
Fig.~\ref{fig:DF_ASIMM_E_PROF} (green dash-dotted line).

Moreover, to study the effects induced by the asymmetry of the peaks on the \ion{H}{I}
outflow velocity, we also took into account the synthetic and reddest asymmetric profile
($I\tiny{peak-blue}/I\tiny{peak-red}=0.86$), which was obtained by taking the opposite
of the bluest asymmetric profile (characterised by the ratio
$I\tiny{peak-blue}/I\tiny{peak-red}=1.16$), which is the one observed
on October 28, 1996 (full disc). This profile is shown in panel ({\it c}) of
Fig.~\ref{fig:DF_ASIMM_E_PROF} (red solid line).

The line profiles with the largest and smallest peak separations are those
observed on April 17, 2009 (polar CH) and in the equatorial CH at the central
meridian on 27-29 November, 1975, respectively. The separation variation is equal
to about 50\% with respect to the smallest separation value. The profiles are
shown in panel ({\it a}) of Fig.~\ref{fig:prof_sep} (red dotted line and green
solid line, respectively).

We also compared two different profiles relative to ARs. The first one, reported
in Fig.~5a of \citet{Fontenla1988} (AR 2363), shows a remarkable reversal, while
the second one, shown in Fig.~12 of \citet{Fontenla1988} (AR 2340), is characterised
by a negligible reversal. In addition, we analysed the characteristic Ly$\alpha$
profile of a faint flare observed in an AR, which is reported in Fig.~1 of \citet{Lemaire1984},
comparing it to the profile of AR 2340.

Furthermore, we considered the analytic expression for the Ly$\alpha$ irradiance
profile proposed by \citet{Auchere2005} and also used by \citet{Dolei2018,Dolei2019},
together with two cases of the parametrised profile reported in \citet{ikl2018}
(hereafter referred to as IKL). The first one consists of a sum of three Gaussian
components, which was introduced to reproduce the mean Ly$\alpha$ observed profile at solar minimum. The other two profiles were constructed by taking into account the sum
of a {\it k}-function, a straight line that simulates the background, and a reversal
Gaussian function reproducing the line reversal. Such profiles were calculated for both
the maximum and minimum irradiance values reported in Table~A.1 of \citet{Lemaire2015}.
These profiles are presented in panel ({\it b}) of Fig.~\ref{fig:profili_osservati_param}
(Auch\`ere: blue solid line; IKL at minimum: green dashed line; IKL at maximum: red dash-dotted line).

In order to calibrate the radiance of the spectral data from SUMER, we considered
the absolute values of line radiance $I \,(\,\mathrm{erg \,cm^{-2} \,s^{-1} \,sr^{-1}}$)
provided by the SOLar STellar Irradiance Comparison Experiment \citep[SOLSTICE;][]{RottmanWoods1994}
aboard the Upper Atmosphere Research Satellite \citep[UARS;][]{Reber1990},
as explained, for example, in \citet{Lemaire1998,Lemaire2002,Lemaire2015}.
The solid angle subtended by the solar disc at the distance of SOLSTICE ($\sim 1$~AU)
was taken into account. The considered profiles were all normalised to the same
value of absolute line radiance observed at the minimum, such as that acquired
on May 22, 1997, which corresponds to the SUMER observation closest in time to the coronal Ly$\alpha$ map reconstructed from UVCS observations reported in this paper.
This value is $I = I_{min} = 8.59\times10^4  \,\mathrm{erg \,cm^{-2} \,s^{-1} \,sr^{-1}}$
\citep[see Table 1 of][]{Lemaire2015}.

\subsection{Parameters for the synthesis of the Ly$\alpha$ scattered intensity}

In order to estimate $v_w$, we synthesised the intensity of the coronal Ly$\alpha$ to be
compared with UVCS observations using the code described in \citet{Dolei2018,Dolei2019}. For the computation of the synthetic coronal emission, we need information about the
physical  quantities on which the Ly$\alpha$  intensity depends, such as $n_e$, $T_e,$
and $T_{\ion{H}{I,}}$ under some assumptions and approximations.

The electron density $n_e$ is obtained through the inversion method developed by
\citet{VanDeHulst1950} that describes the polarised brightness (due to the Thomson
scattering) as dependent only on the electron density, under the hypothesis
of cylindrical symmetry \citep[see, e.g.,][]{Hayes2001,Dolei2015}. We used the
profile of $n_e$ derived from the LASCO {\it pB} map, as described in \citet{Dolei2018}.

For the electron temperature $T_e,$ we used the values derived by \citet{Dolei2019}.
They obtained $T_e$ as a function of the heliocentric distance and latitude, taking
into account the dependence found by \citet{Gibson1999} for equatorial regions and by
\citet{Vasquez2003}
for high latitude regions, under the hypothesis of super-radial expansion of the solar
wind in polar CHs during the solar minimum. Interpolated values for mid
latitudes have been used.

The \ion{H}{I} temperature comes from the database created by \citet{Dolei2016}.
We considered a $T_{\ion{H}{I}}$ map for a solar minimum epoch, following \citet{Dolei2019}.
In order to fit the neutral hydrogen temperature profile and to create a 2D map as a
function of latitude and heliocentric distance, the functional form given
by \citet{Vasquez2003} was used, as in \citet{Dolei2016}.
 It is worth mentioning that the values of
 $T_{\ion{H}{I}}$ are likely overestimated, because the
 hydrogen line broadening includes contributions due to non-thermal mechanisms.
 
For the sake of simplicity, we adopted the approximation $\theta =0^\circ$,
that is, ${\bf{v}}\parallel{\mathbf{n}^{\prime}}$ in all the computations
leading to the results  described in Sect.~\ref{sec:res}. However, the consequence of
assuming values of $\theta \ne 0$ on the inferred values of the \ion{H}{I} outflow
velocities has been evaluated as a separate step and is also described in Sect.~\ref{sec:res}.

\section{Results}
\label{sec:res}

\begin{figure}[h!]
        \centering
        \includegraphics[trim=0 90 0 20, clip, width=\hsize]{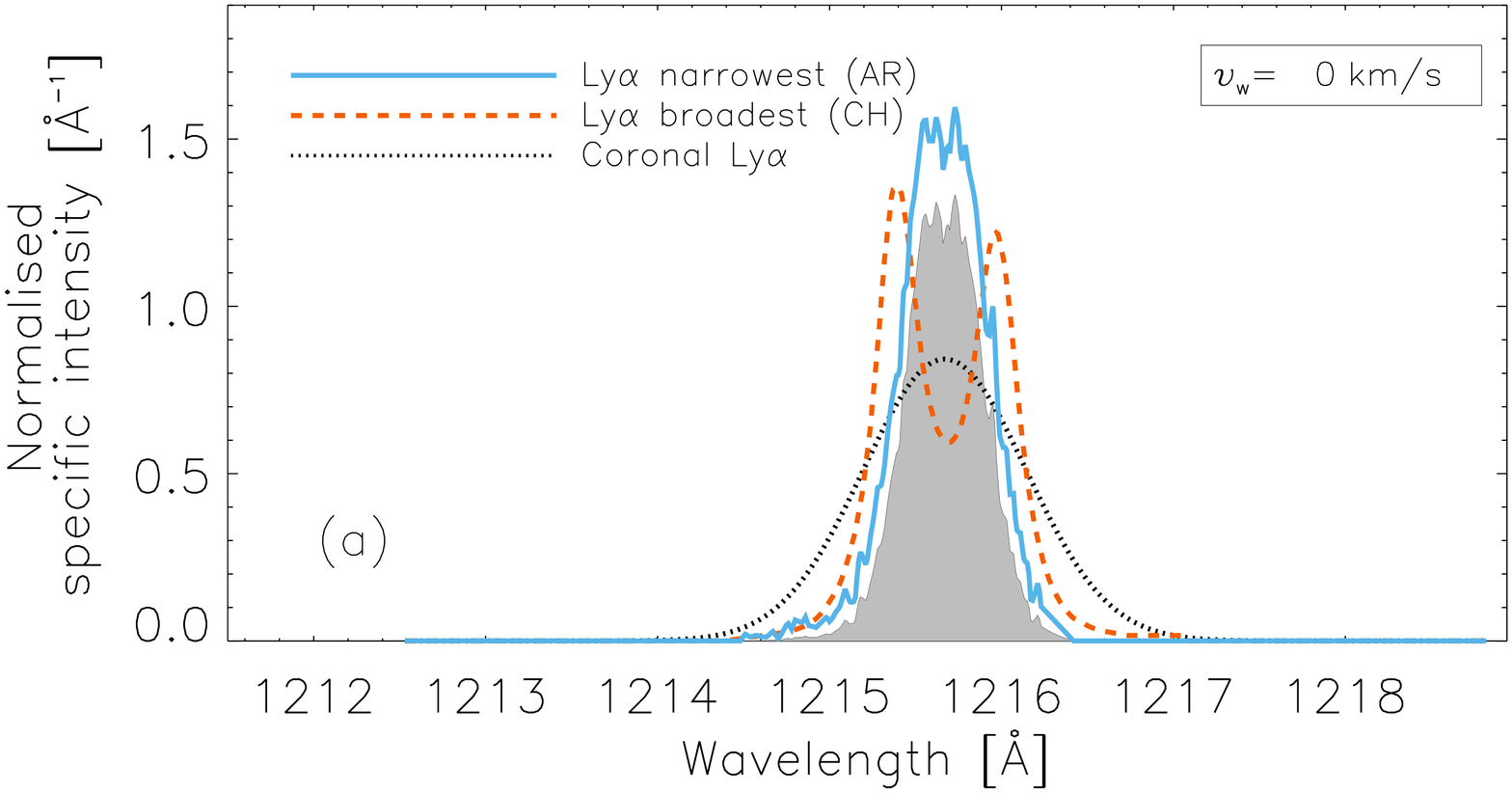}
        \includegraphics[trim=0 90 0 20, clip, width=\hsize]{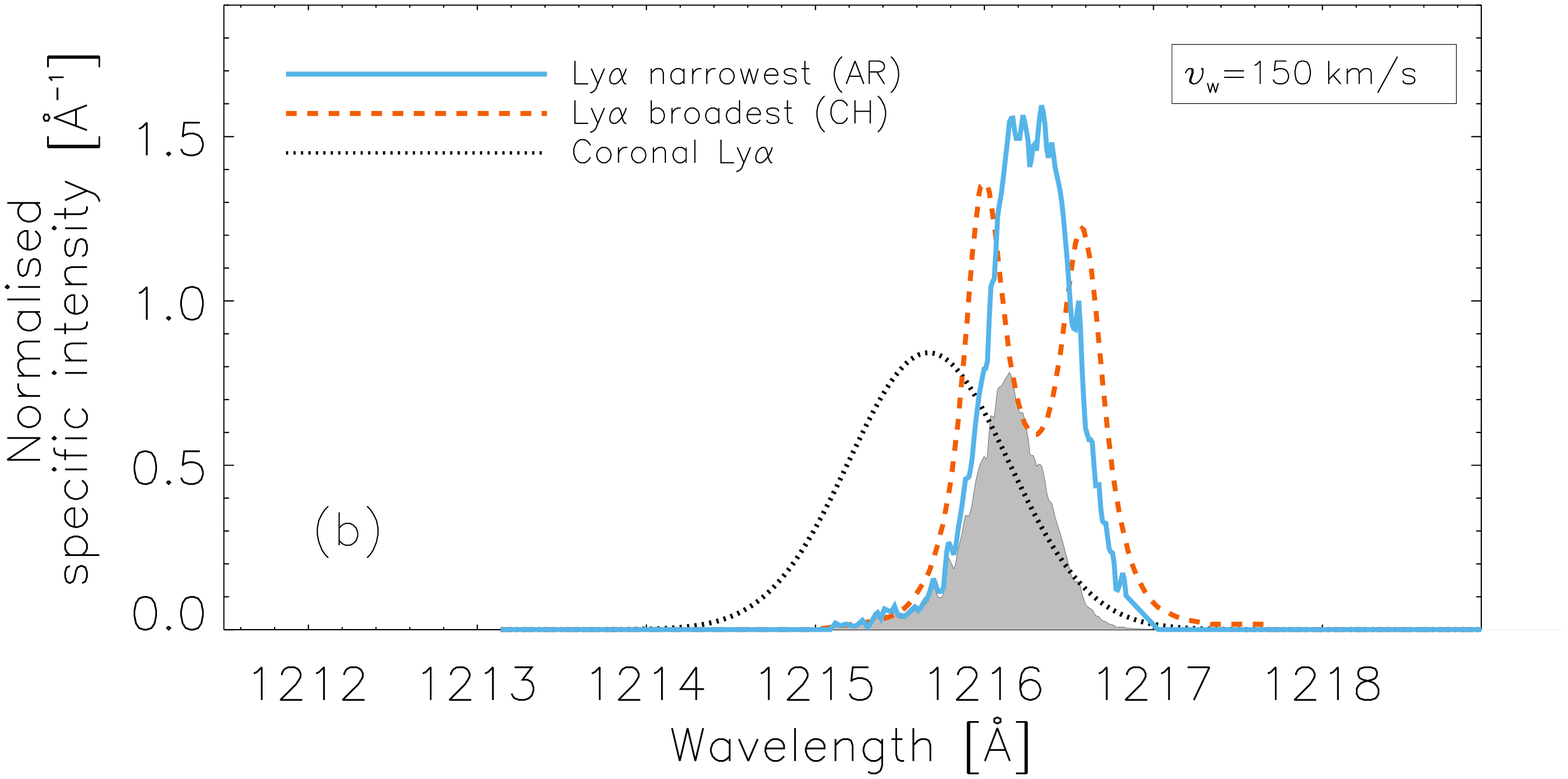}
        \includegraphics[trim=0 0 0 20, clip, width=\hsize]{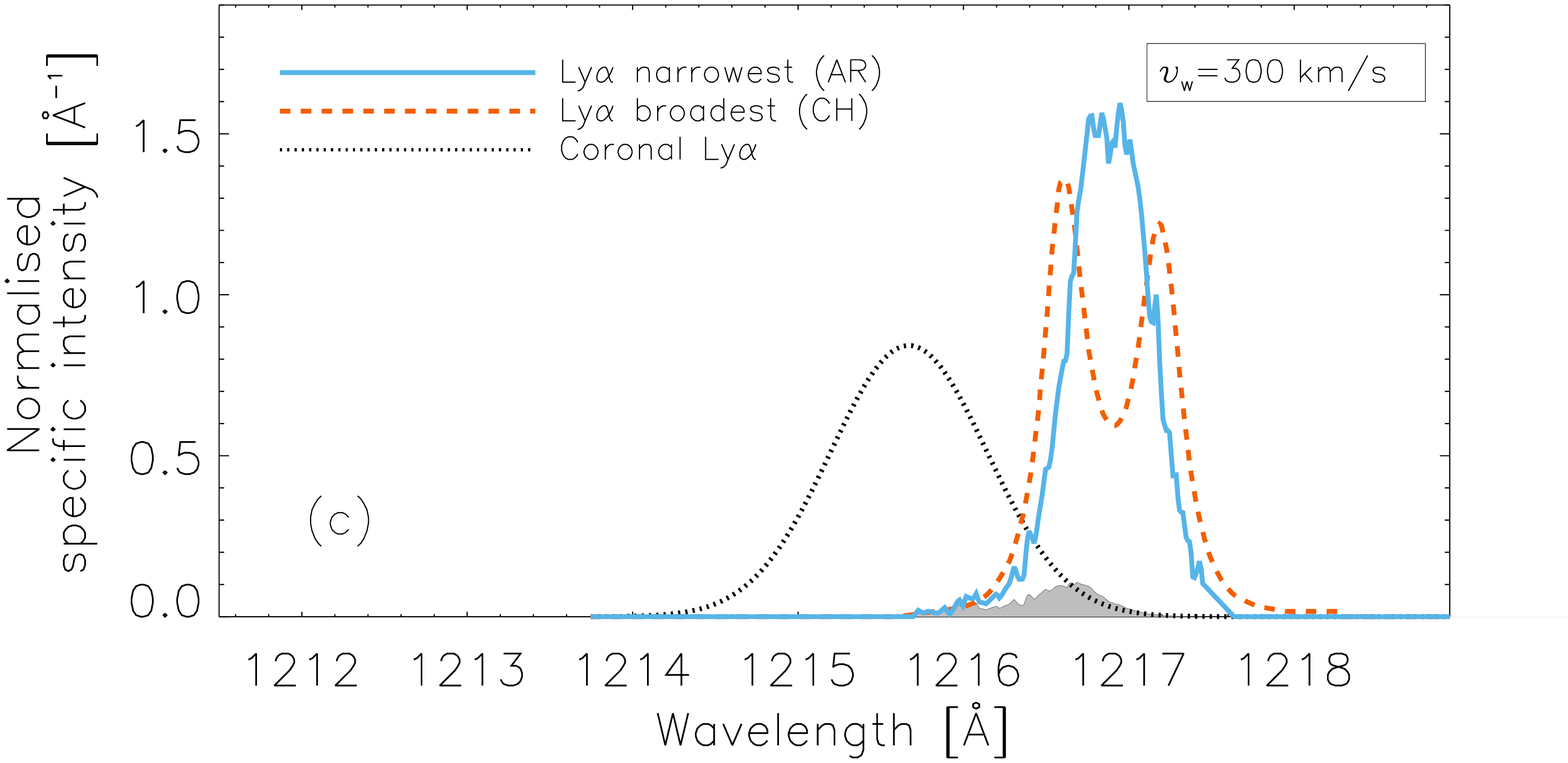}
        \includegraphics[trim=0 0 0 20, clip, width=\hsize]{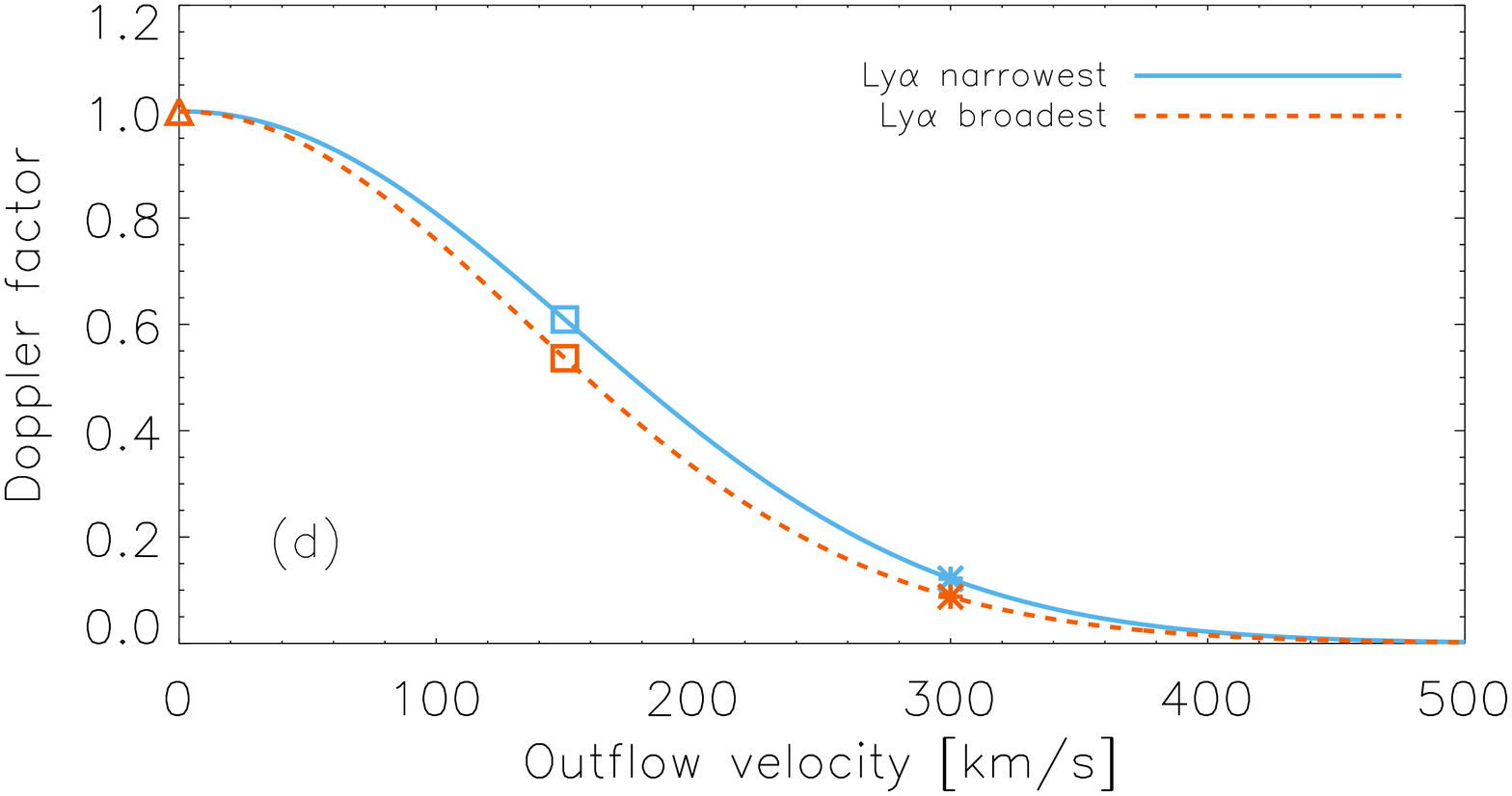}
        \caption{
        Panels ({\it a}), ({\it b}), and ({\it c}): Normalised chromospheric profiles reported
        in Fig.~12 of \citet{Fontenla1988} (AR 2340, blue solid line) and observed on
        April 17, 2009 \citep[red dashed line;][]{Tian2009b}, overlapped with a normalised
        Gaussian coronal absorption profile (black dotted line) computed by setting
        the coronal \ion{H}{I} temperature to $1.5\times10^6$~K, considering $\theta = 0^\circ$.
        Panels ({\it a}), ({\it b}), and ({\it c}) correspond to outflow velocities
        equal to $v_w =0  \,\mathrm{km \,s}^{-1}$, $v_w$~=$~150  \,\mathrm{km \,s}^{-1}$
        , and $v_w = 300  \,\mathrm{km \,s}^{-1}$, respectively. In the case of the exciting
        narrowest profile, we also display the grey shaded area that is proportional
        to the product of the coronal absorption profile ($\Phi$) and the chromospheric
        pumping profile ($\Psi$). Panel ({\it d}): Doppler factor as a function of $v_w$
        calculated considering each chromospheric profile (AR 2340: blue solid line; polar CH:
        red dashed line). Triangles, squares and asterisks indicate the Doppler dimming
        values corresponding to \ion{H}{I} outflow velocities equal to
        $v_w =0  \,\mathrm{km \,s}^{-1}$, $v_w$~=$~150  \,\mathrm{km \,s}^{-1}$
        , and $v_w = 300  \,\mathrm{km \,s}^{-1}$, respectively.
                }
        \label{fig:DF}
\end{figure}

In order to verify the effects of the chromospheric Ly$\alpha$ profile shape on the
determination of the solar wind \ion{H}{I} velocity, we first studied the
behaviour of the Doppler factor dependent on the profile width. The results
are summarised in Fig.~\ref{fig:DF}. The first three panels show the overlapping
between the narrowest and broadest chromospheric profiles and the coronal one that
has been calculated by assuming a \ion{H}{I} temperature in the corona equal to
$1.5\times10^6$~K. We took into account the following velocity values:
$v_w = 0  \,\mathrm{km \,s}^{-1}$ (panel ({\it a})), $v_w = 150  \,\mathrm{km \,s}^{-1}$
(panel ({\it b})), and $v_w = 300  \,\mathrm{km \,s}^{-1}$ (panel ({\it c})).
In addition, for the case of the narrowest profile, we also display the grey shaded
area that is proportional to the integrand value of the function $F({\bf{n'}},v_w, \theta)$
(see Eq.~\ref{eq:equaz_doppler_factor}). This quantity is closely related to the Doppler
factor as it gives the overlapping between the coronal absorption profile ($\Phi$)
and the chromospheric pumping profile ($\Psi$).

In Fig.~\ref{fig:DF}, panel ({\it d}), the corresponding values of $D(v_w)$
are properly labelled on the curves, where the Doppler factor as a function of $v_w$
(varying from $v_w = 0  \,\mathrm{km \,s}^{-1}$ to $v_w = 500  \,\mathrm{km \,s}^{-1}$)
is reported. We note that the curves in the bottom panel show a small difference
between them, indicating that the Doppler dimming does not have a strong dependence
on the variation of the line width of the selected chromospheric profiles.

\begin{figure*}[t]
        \centering
        \includegraphics[width=.5\hsize]{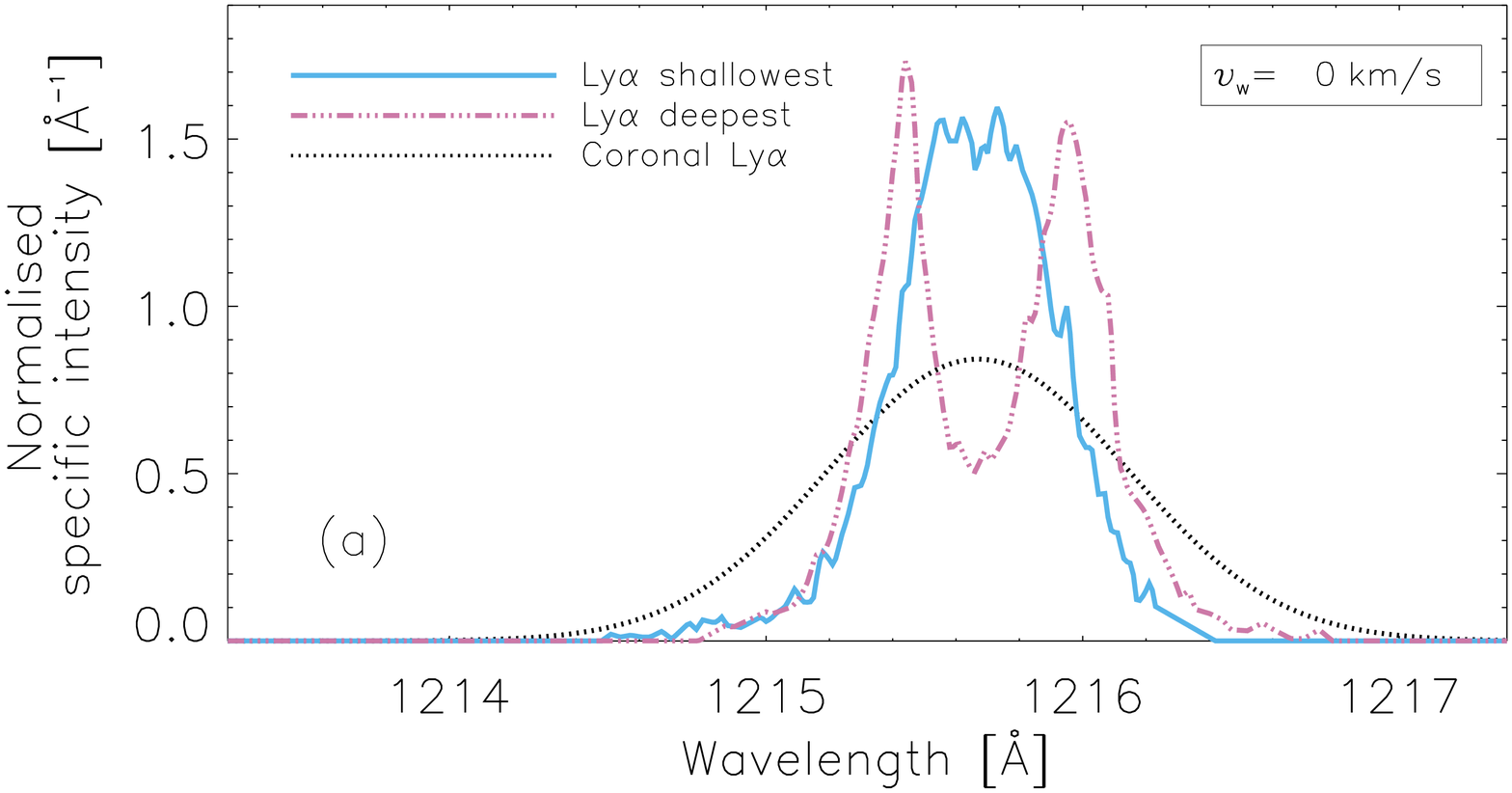}%
        \includegraphics[width=.5\hsize]{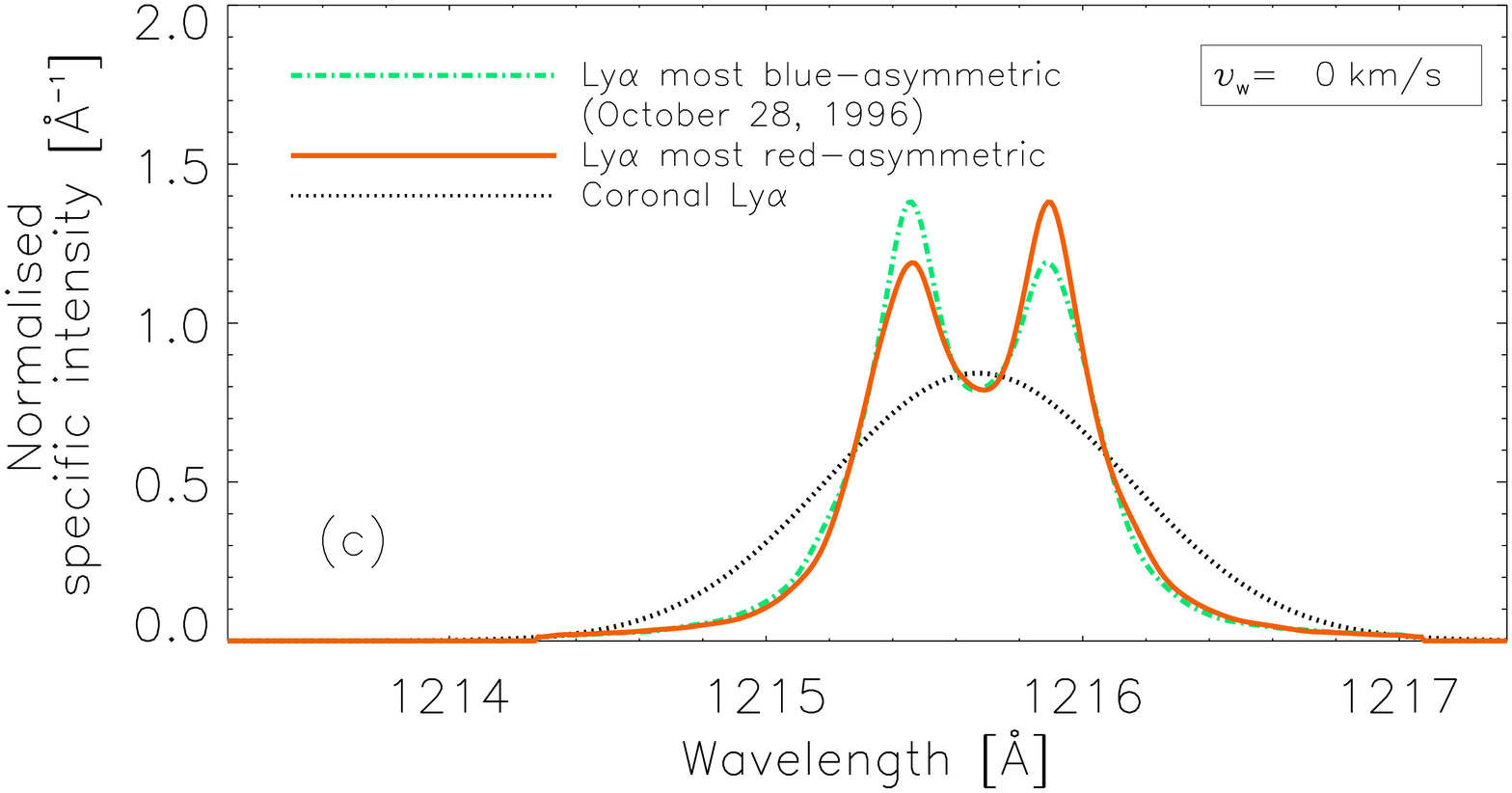}
    \includegraphics[width=.5\hsize]{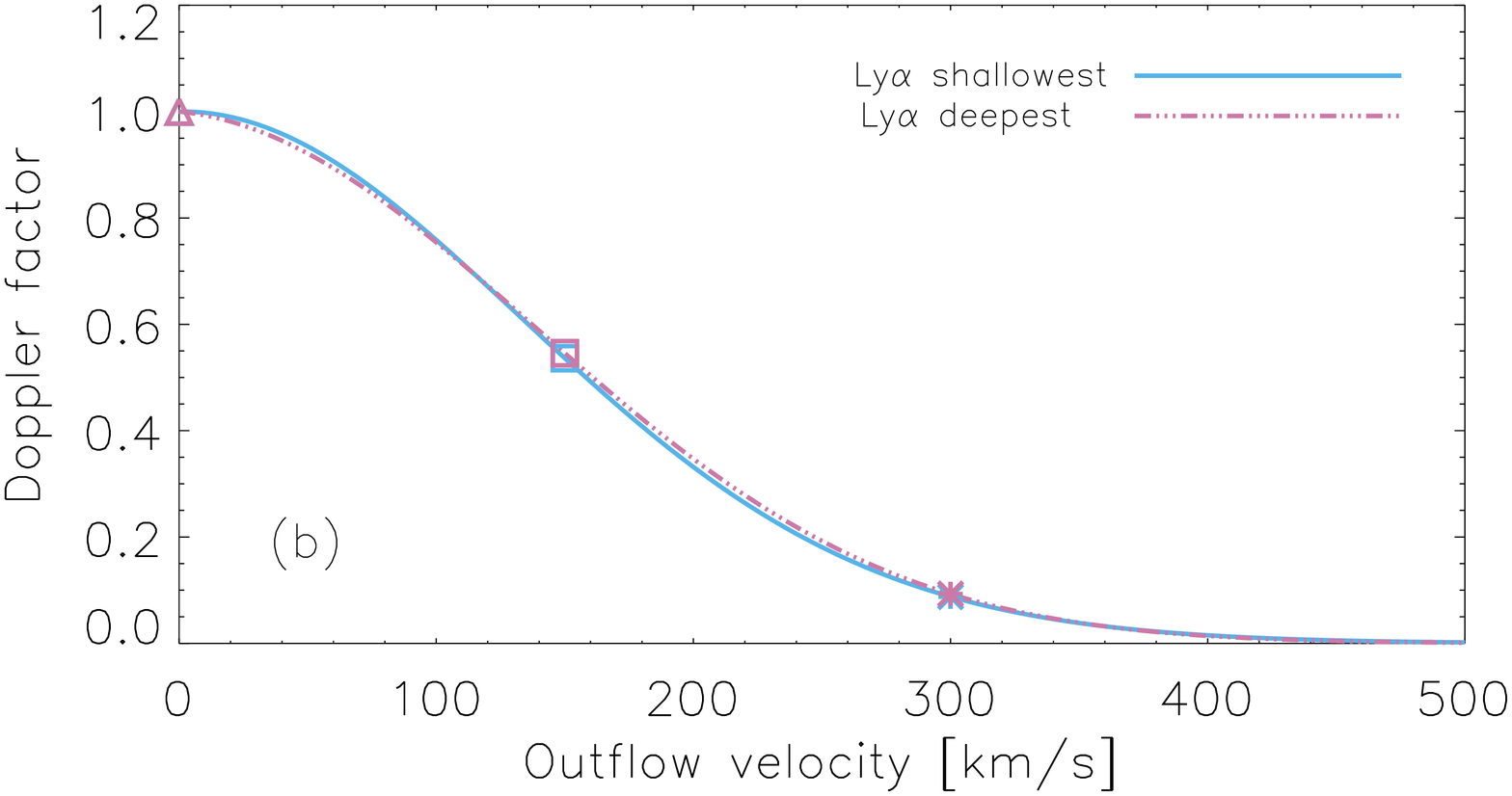}%
    \includegraphics[width=.5\hsize]{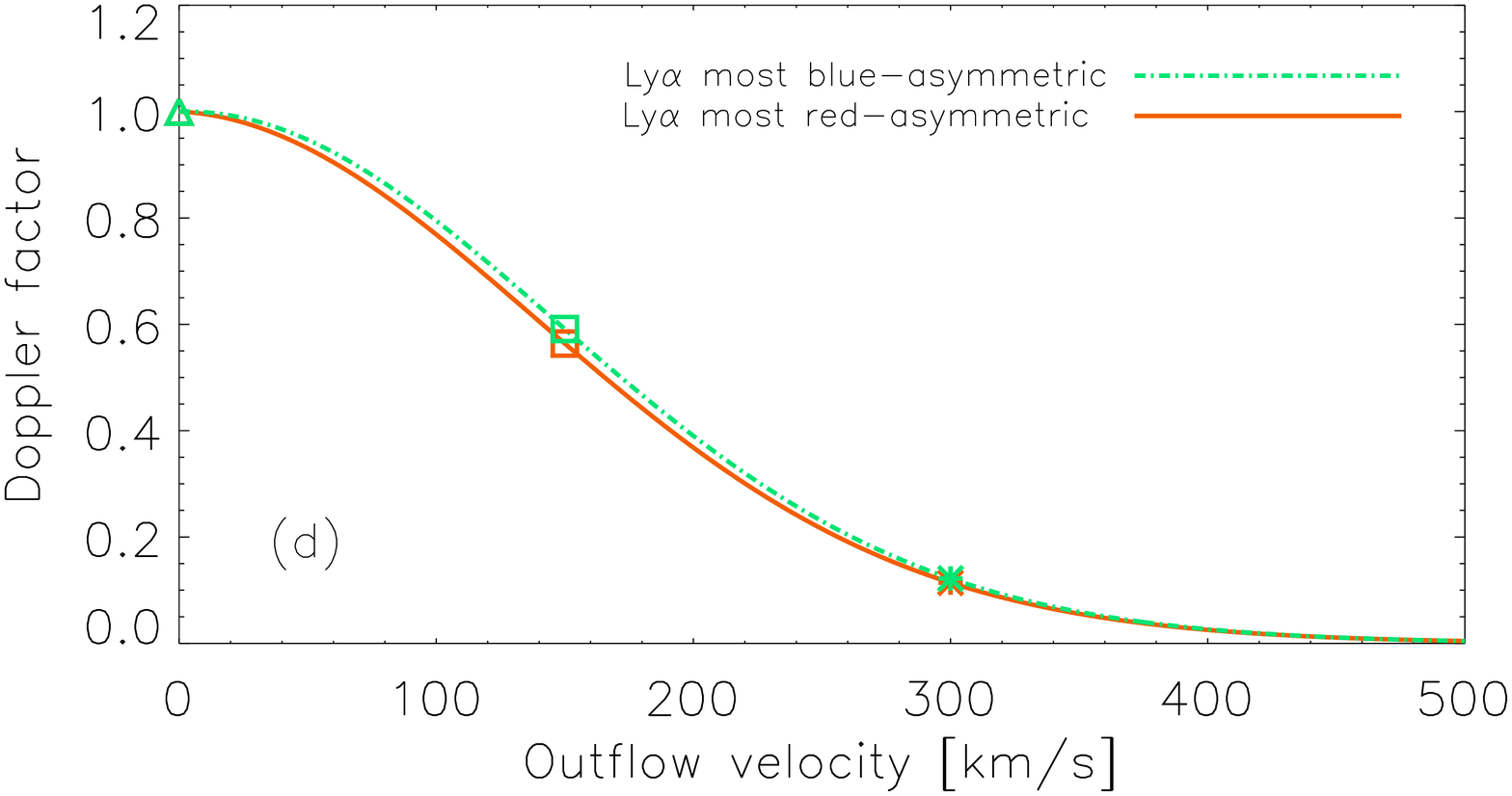}
    \caption{   
        Panel ({\it a}):
        Normalised shallowest (blue solid line) and deepest (magenta dash-dot-dot-dot line)
        profiles described in Fig.~12 and Fig.~6a of \citet{Fontenla1988}, respectively,
        overlapped with a normalised Gaussian coronal profile (black dotted line) computed
        by setting the coronal \ion{H}{I} temperature at $1.5\times10^6$~K, considering
        $\theta = 0^\circ$. Panel ({\it b}): Doppler factor as a function of
        $v_w$ calculated considering each chromospheric profile. Triangles, squares, and
        asterisks: See Fig.~\ref{fig:DF}.
        Panels ({\it c}) and ({\it d}): Same as in panels ({\it a}) and ({\it b}),
        but concerning the bluest asymmetric profile ($I\tiny{peak-blue}/I\tiny{peak-red}=1.16$)
        observed on October 28, 1996 \citep[green dash-dotted line;][]{Lemaire2015} and
        the reddest asymmetric profile ($I\tiny{peak-blue}/I\tiny{peak-red}=0.86$;
        red solid line), respectively.
        }
        \label{fig:DF_ASIMM_E_PROF}
\end{figure*}

\begin{figure}[t]
        \centering
        \includegraphics[width=\hsize]{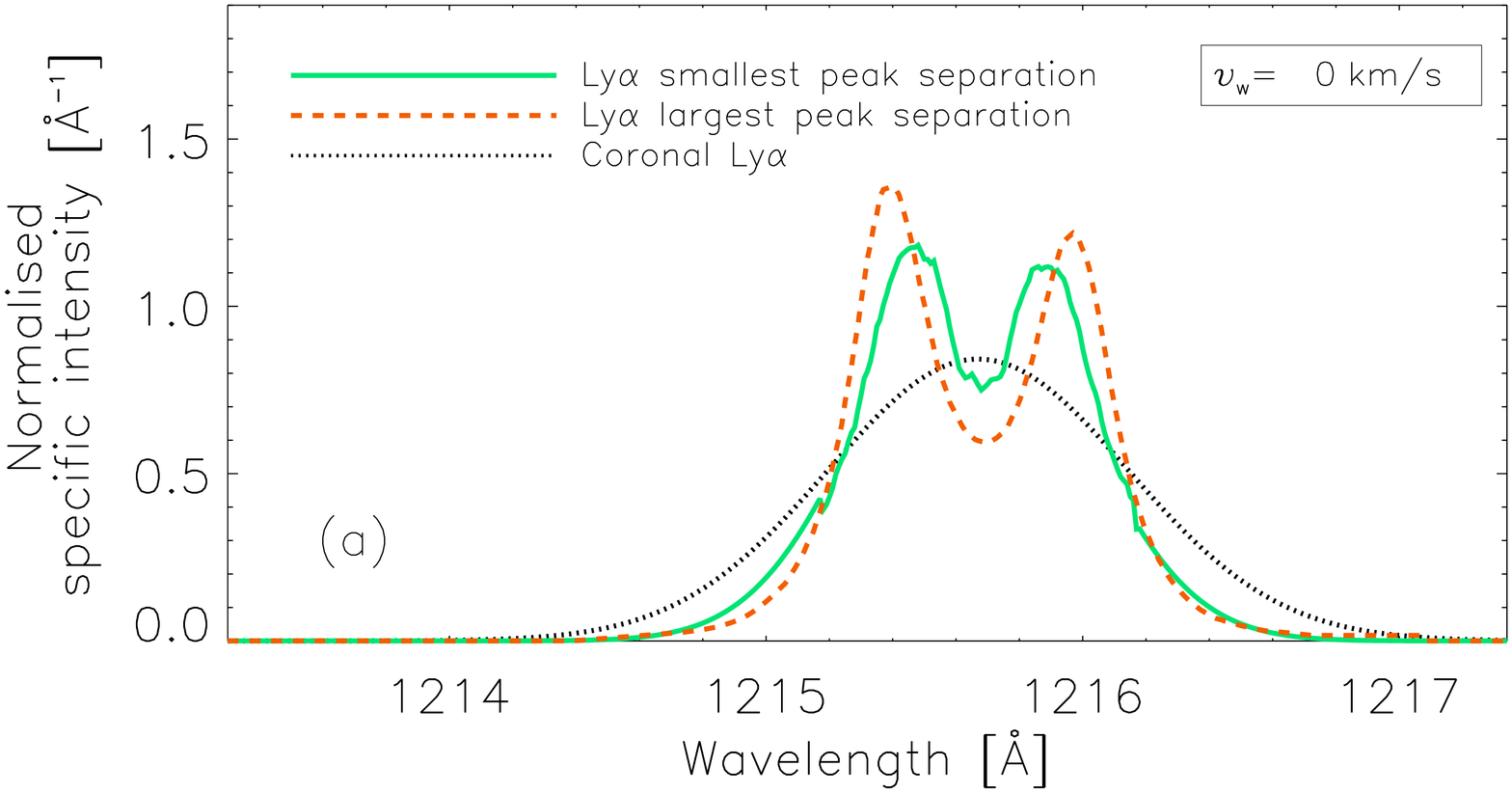}
       \includegraphics[width=\hsize]{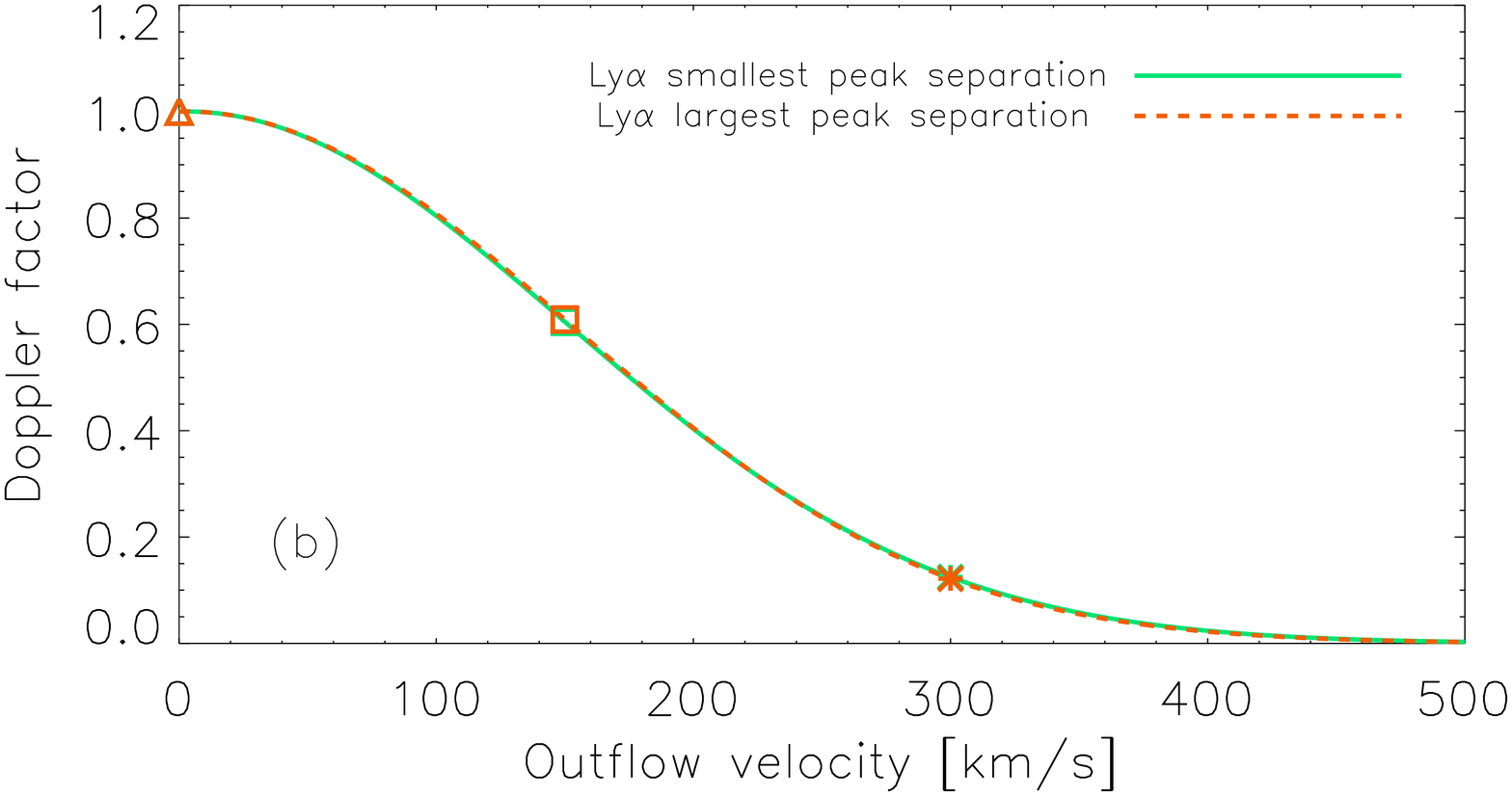}
        \caption{
        Panel ({\it a}):
        Normalised CH profiles with smallest (green solid line) and greatest
        (red dashed dotted line) peak separations observed in an equatorial CH
        at the central meridian on 27-29 November, 1975 \citep{BocchialiniVial1996},
        and on April 17, 2009 \citep[polar CH;][]{Tian2009b}, respectively,
        overlapped with a normalised Gaussian coronal profile (black dotted line)
        computed by setting the coronal \ion{H}{I} temperature at $1.5\times10^6$~K,
        considering $\theta = 0^\circ$. Panel ({\it b}): Doppler factor as a function
        of $v_w$ calculated considering each chromospheric profile. Triangles, squares,
        and asterisks: See Fig.~\ref{fig:DF}.
        }
        \label{fig:prof_sep}
\end{figure}

We studied the Doppler factor $D(v_w)$ as a function of the profile reversal depth.
The results are reported in Fig.~\ref{fig:DF_ASIMM_E_PROF}.
Panel ({\it a}) of Fig.~\ref{fig:DF_ASIMM_E_PROF} shows the overlapping of the coronal
absorption profile with the shallowest and the deepest chromospheric profiles. The same
analysis was carried out for the bluest and reddest asymmetric profiles (panel ({\it c})
of Fig.~\ref{fig:DF_ASIMM_E_PROF}) and the profiles with the smallest and the largest
separation between the peaks (panel ({\it a}) of Fig.~\ref{fig:prof_sep}).
The curves in panel ({\it b}) and ({\it d}) of Fig.~\ref{fig:DF_ASIMM_E_PROF} and panel
({\it b}) of Fig.~\ref{fig:prof_sep} do not show any significant differences between them.
Therefore, the Doppler factor does not have a
strong dependence on reversal depth, asymmetry, and separation of the peaks with regard to all the selected chromospheric profiles taken into account as extreme cases.

\begin{figure}[t]
        \centering
        \includegraphics[width=\hsize]{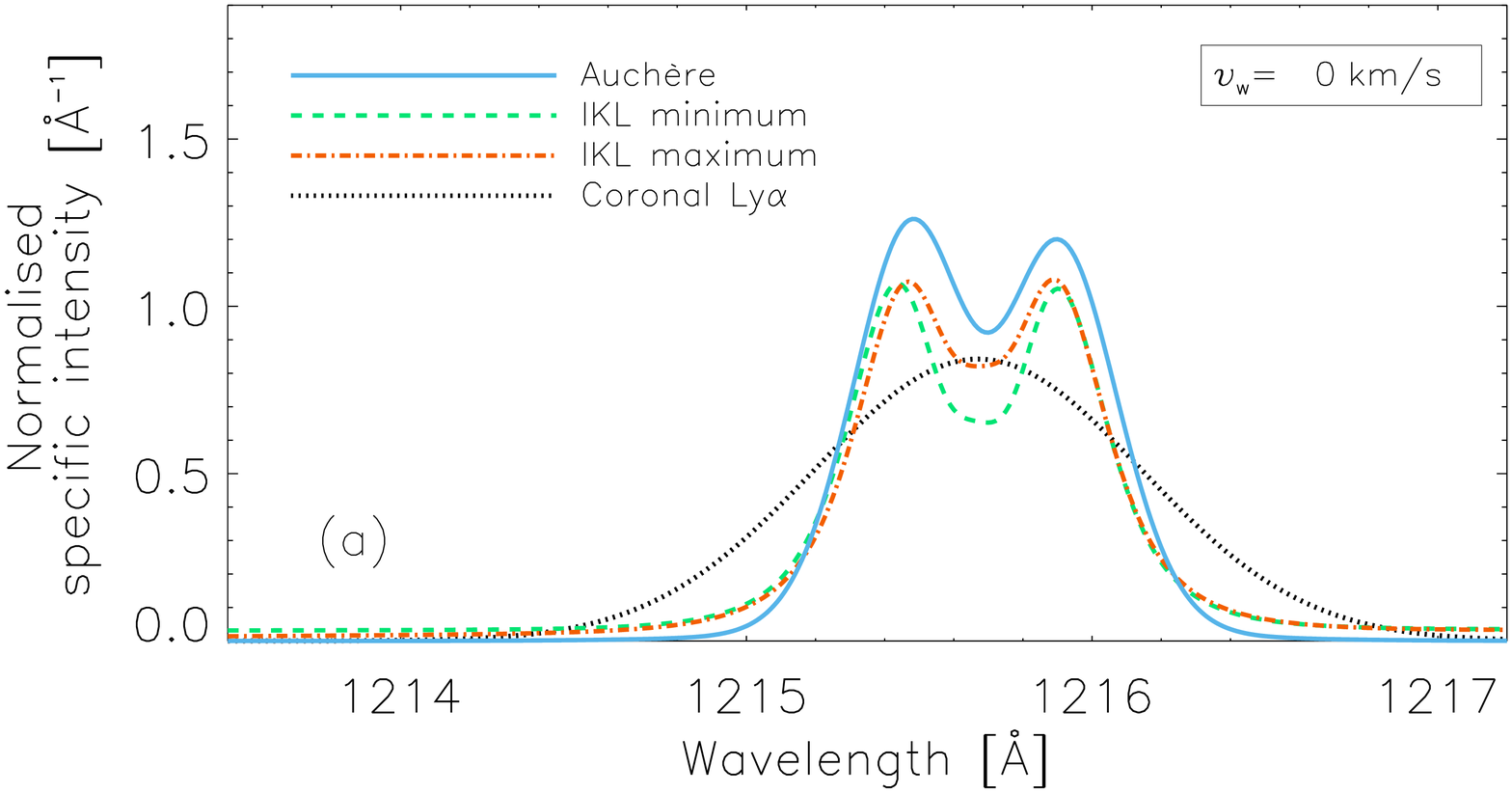}
        \includegraphics[width=\hsize]{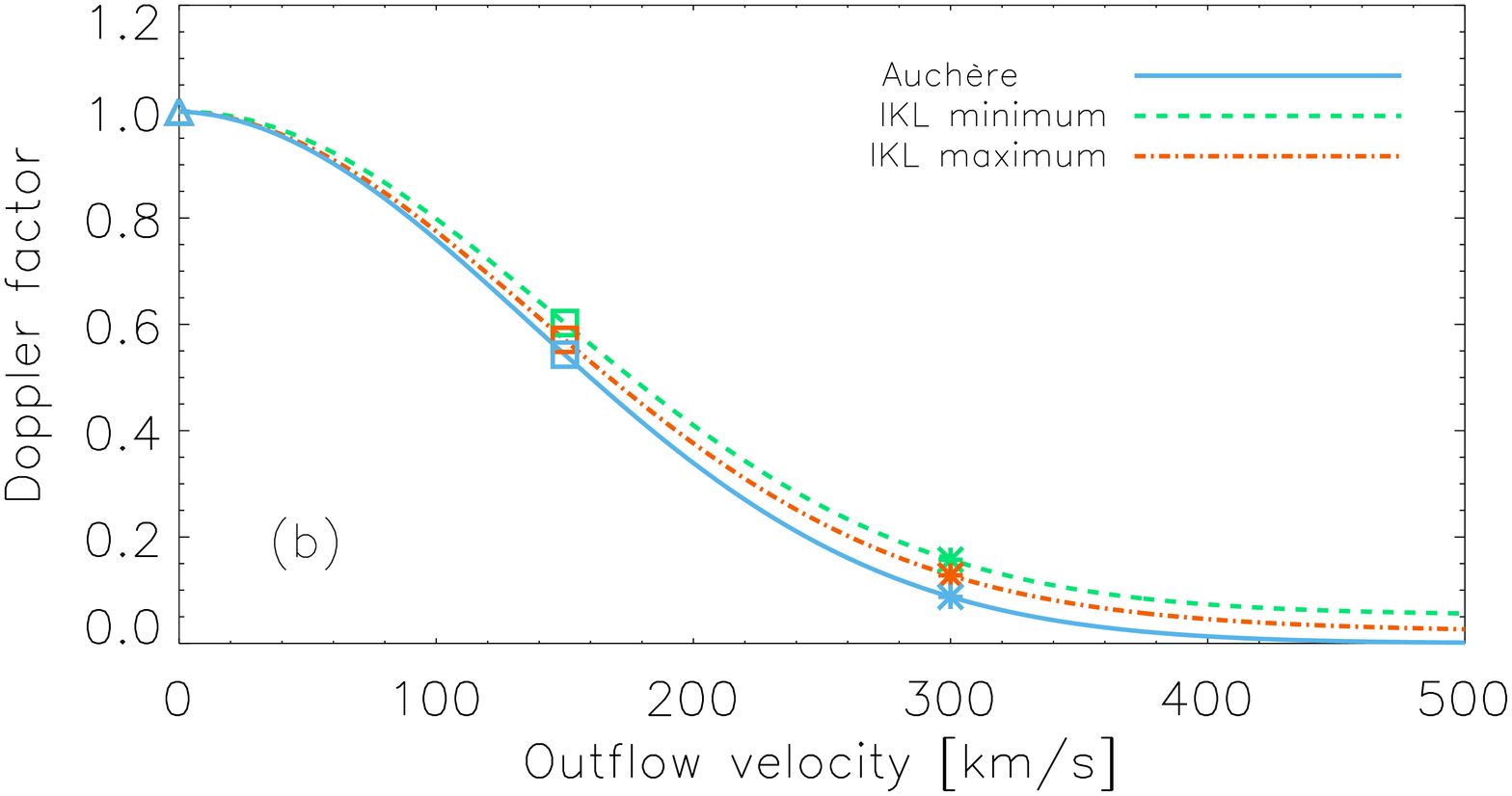}      
        \caption{
        Panel ({\it a}):
        Normalised parametrised chromospheric profiles reported in
        Fig.~\ref{fig:profili_osservati_param}; \citet{Auchere2005}
        (blue solid line) and \citet{ikl2018} (green dashed line at
        minimum and red dot-dashed line at maximum), overlapped
        with a normalised Gaussian coronal profile (black dotted line)
        computed by setting the coronal \ion{H}{I} temperature at
        $1.5\times10^6$~K, considering $\theta = 0^\circ$.
        Panel ({\it b}): Doppler factor relative to the profiles shown
        in panel ({\it a}). Triangles, squares, and asterisks: See Fig.~\ref{fig:DF}.
        }
        \label{fig:DF_prof_parametrizzati}
\end{figure}

We performed a similar analysis by considering the three profiles reported in
Fig.~\ref{fig:profili_osservati_param} (panel ({\it b})).
From Fig.~\ref{fig:DF_prof_parametrizzati} (panels ({\it a}) and ({\it b})), we infer
that the different parametrisations of these chromospheric profiles do not generate
significant differences in the Doppler factor values, even if they are larger than in
the observed profiles.

\begin{figure*}[t]
        \centering
        \includegraphics[width=.5\hsize]{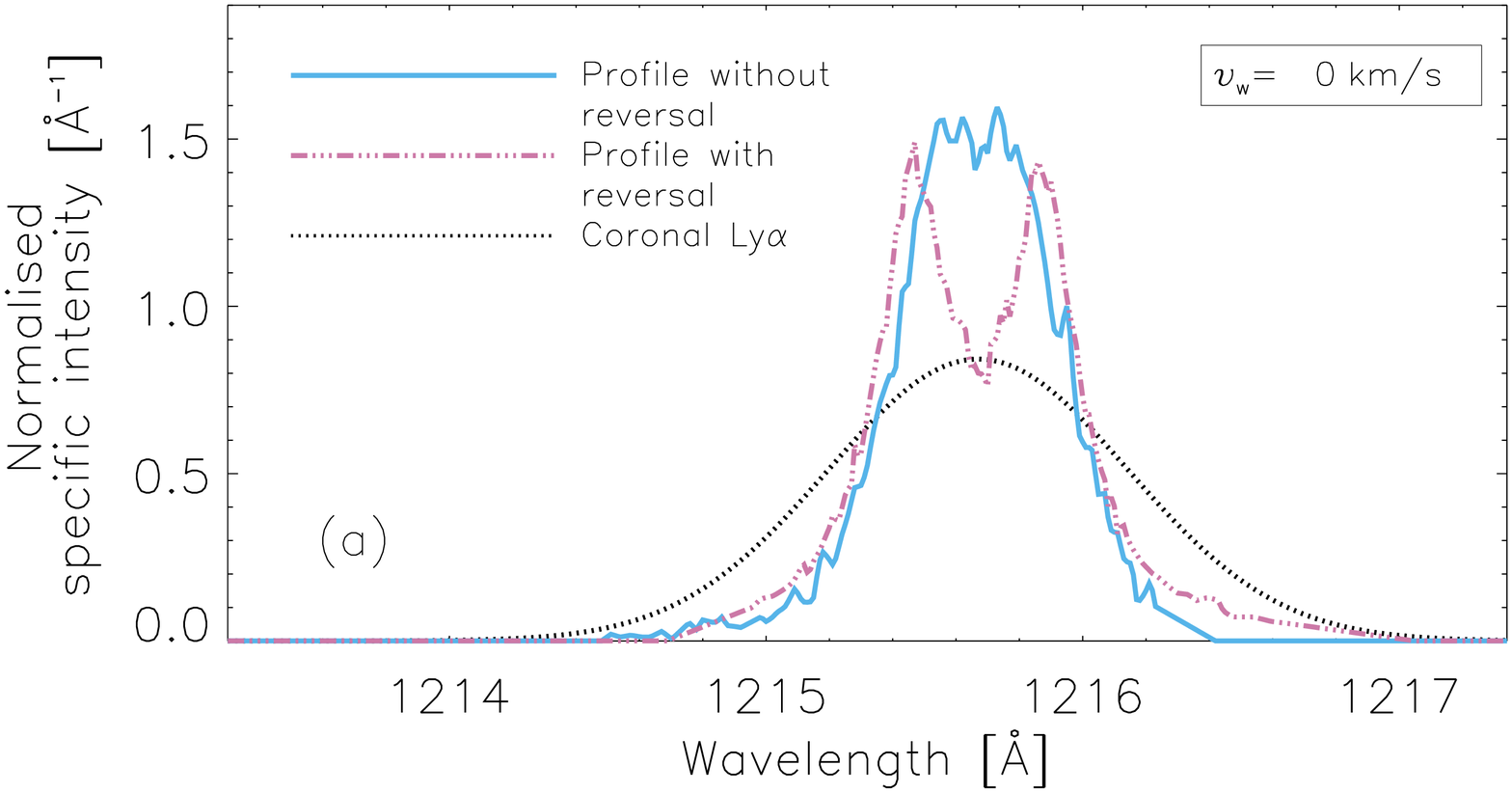}%
        \includegraphics[width=.5\hsize]{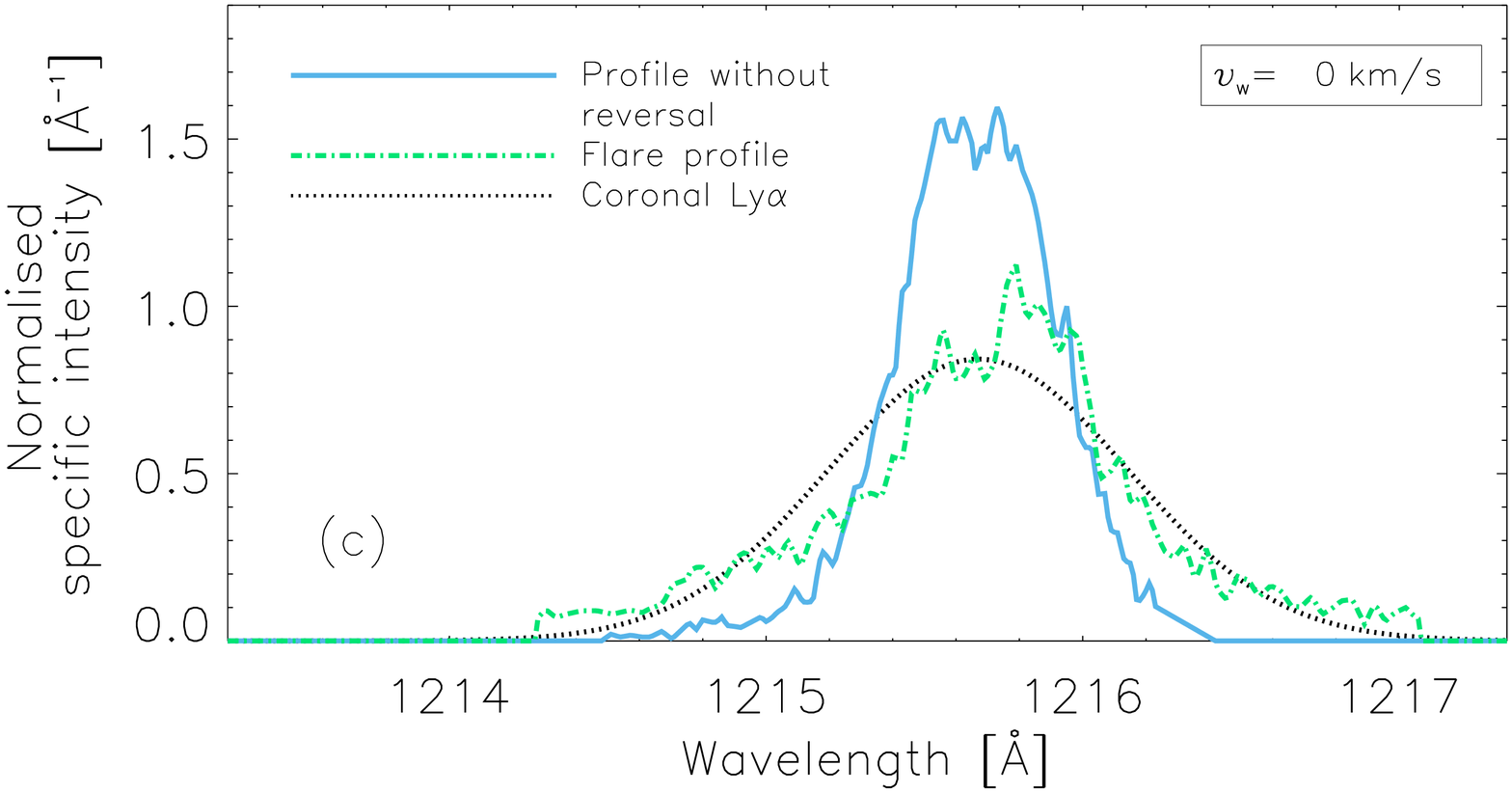}
        \includegraphics[width=.5\hsize]{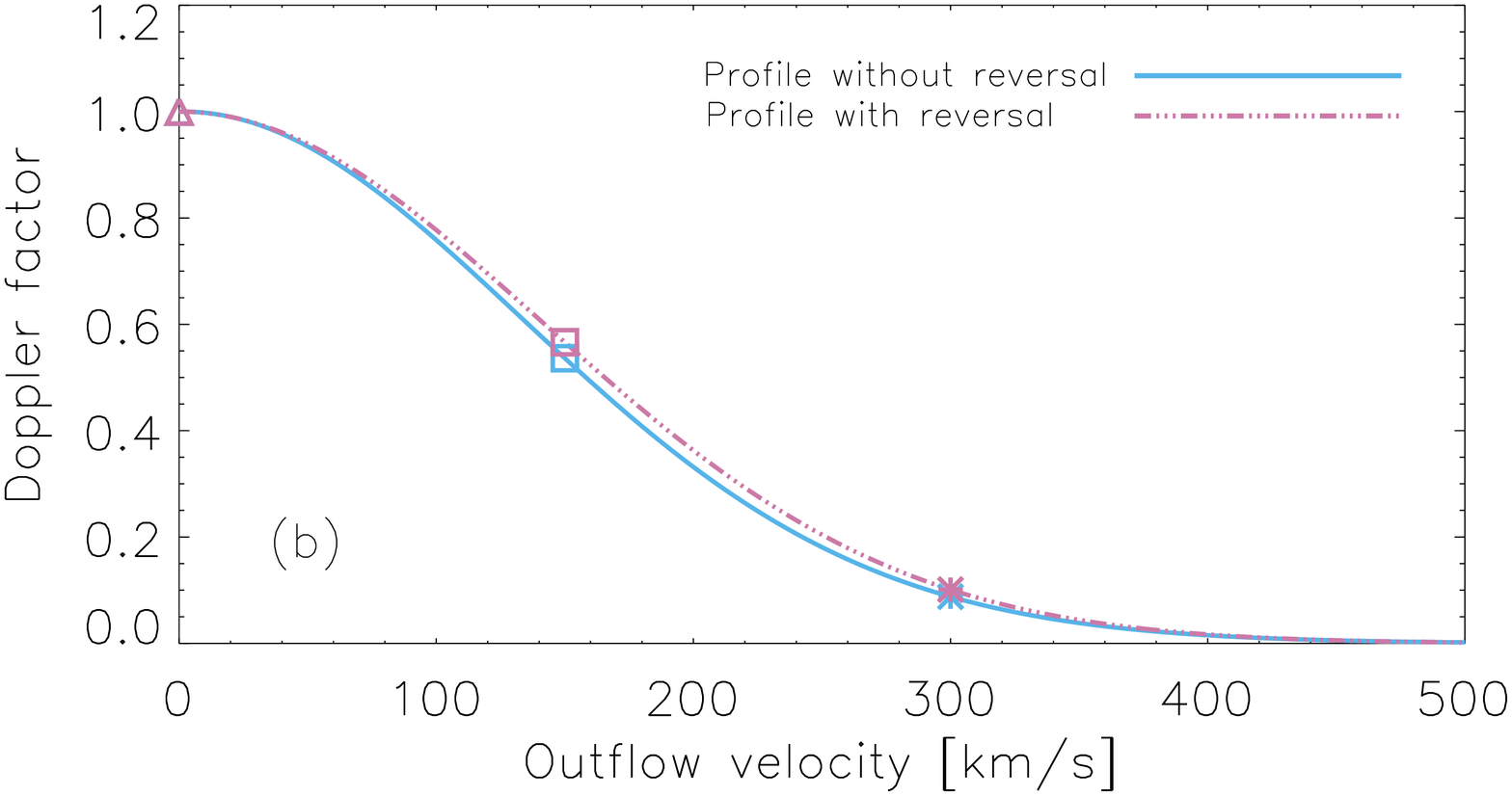}%
        \includegraphics[width=.5\hsize]{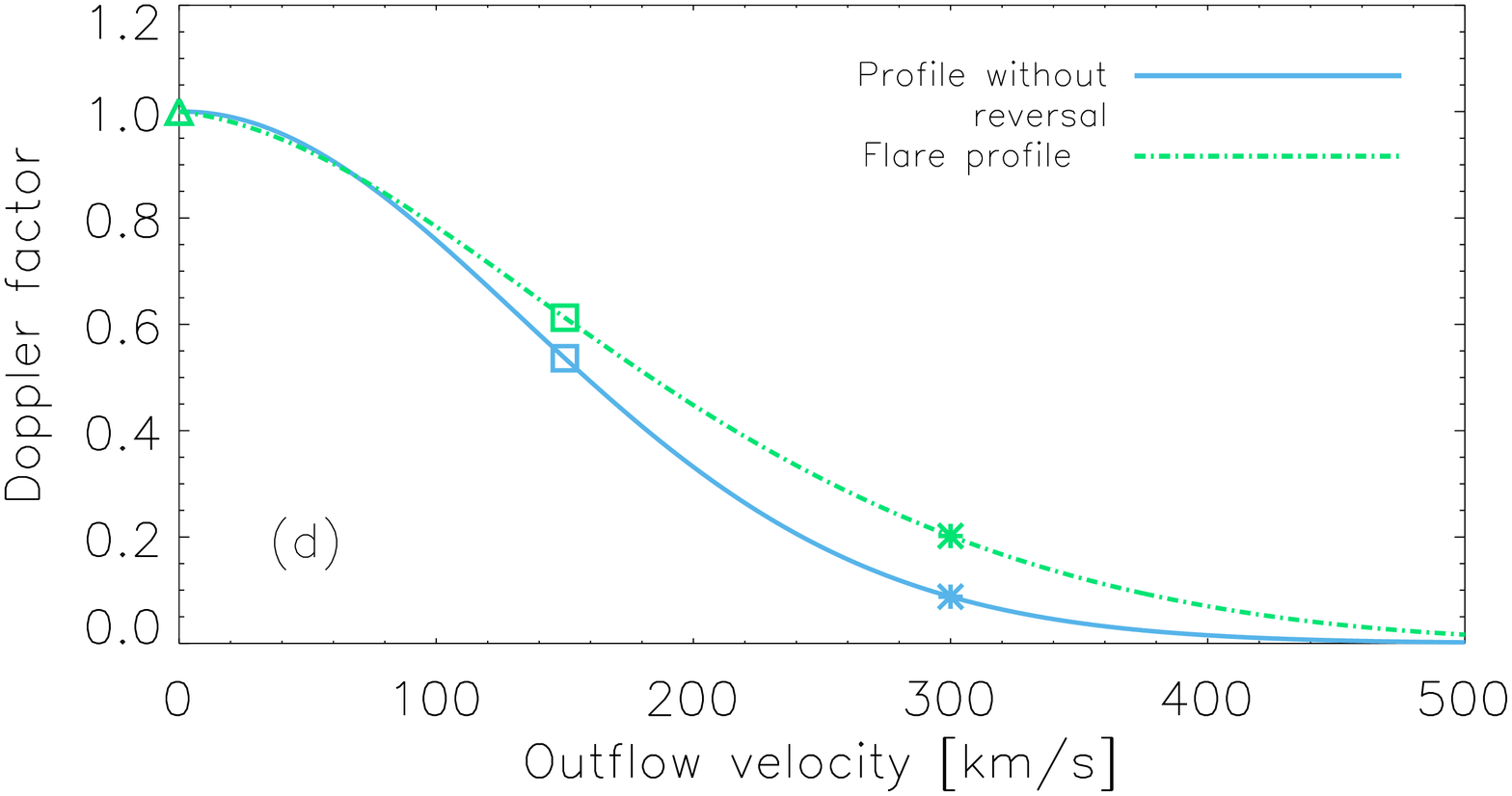}
        \caption{
        Panel ({\it a}): Normalised AR profiles with and without reversal relevant
        to AR 2363 (magenta dash-dot-dot-dot line) and AR 2340 (blue solid line)
        \citep[see][]{Fontenla1988}, respectively, overlapped with a normalised
        Gaussian coronal profile (black dotted line) computed by setting the coronal
        \ion{H}{I} temperature at $1.5\times10^6$~K, considering $\theta = 0^\circ$.
        Panel ({\it b}): Doppler factor as a function of $v_w$ calculated considering
        each chromospheric profile. Triangles, squares, and asterisks: See Fig.~\ref{fig:DF}.
        Panels ({\it c}) and ({\it d}): Same as in panels ({\it a}) and ({\it b}), but
        concerning the AR 2340 profile \citep[blue solid line;][]{Fontenla1988} and
        the normalised flare profile \citep[green dash-dotted line;][]{Lemaire1984}, respectively.
        }
        \label{fig:ass_senz_ass_e_flare}
\end{figure*}

A further analysis has been carried out for AR profiles. As shown in
Fig.~\ref{fig:ass_senz_ass_e_flare}, when we compare the values of $D(v_w)$
for the AR 2363 (AR showing line reversal) and AR 2340 (AR with a negligible
reversal), we find very small differences. Conversely, comparing the values
relevant to the AR 2340 and the flare profiles, we observe significant differences.

\begin{figure}[t]
        \centering
        \includegraphics[width=\hsize]{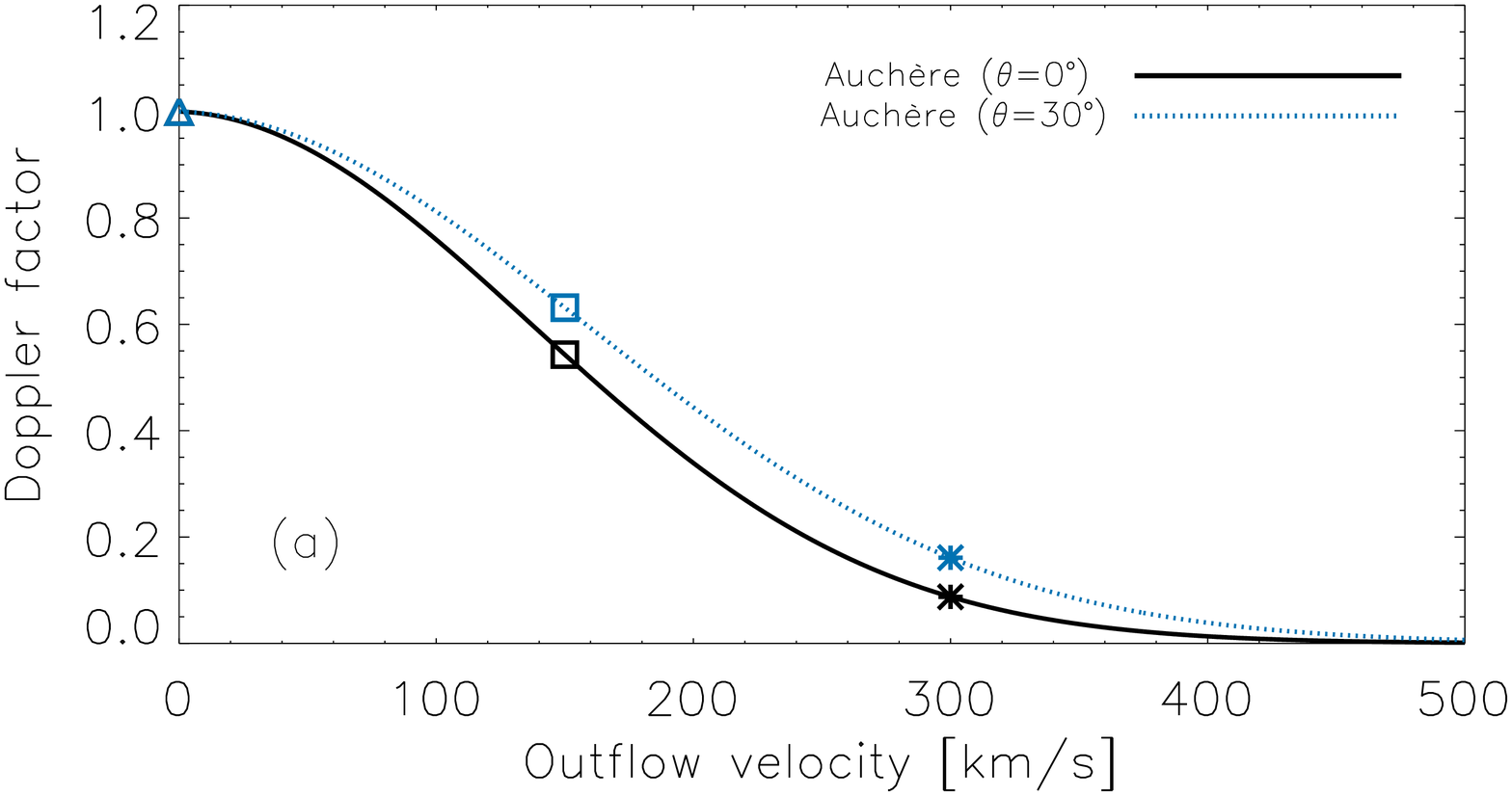}
        \includegraphics[trim=0 0 0 33, clip, width=\hsize]{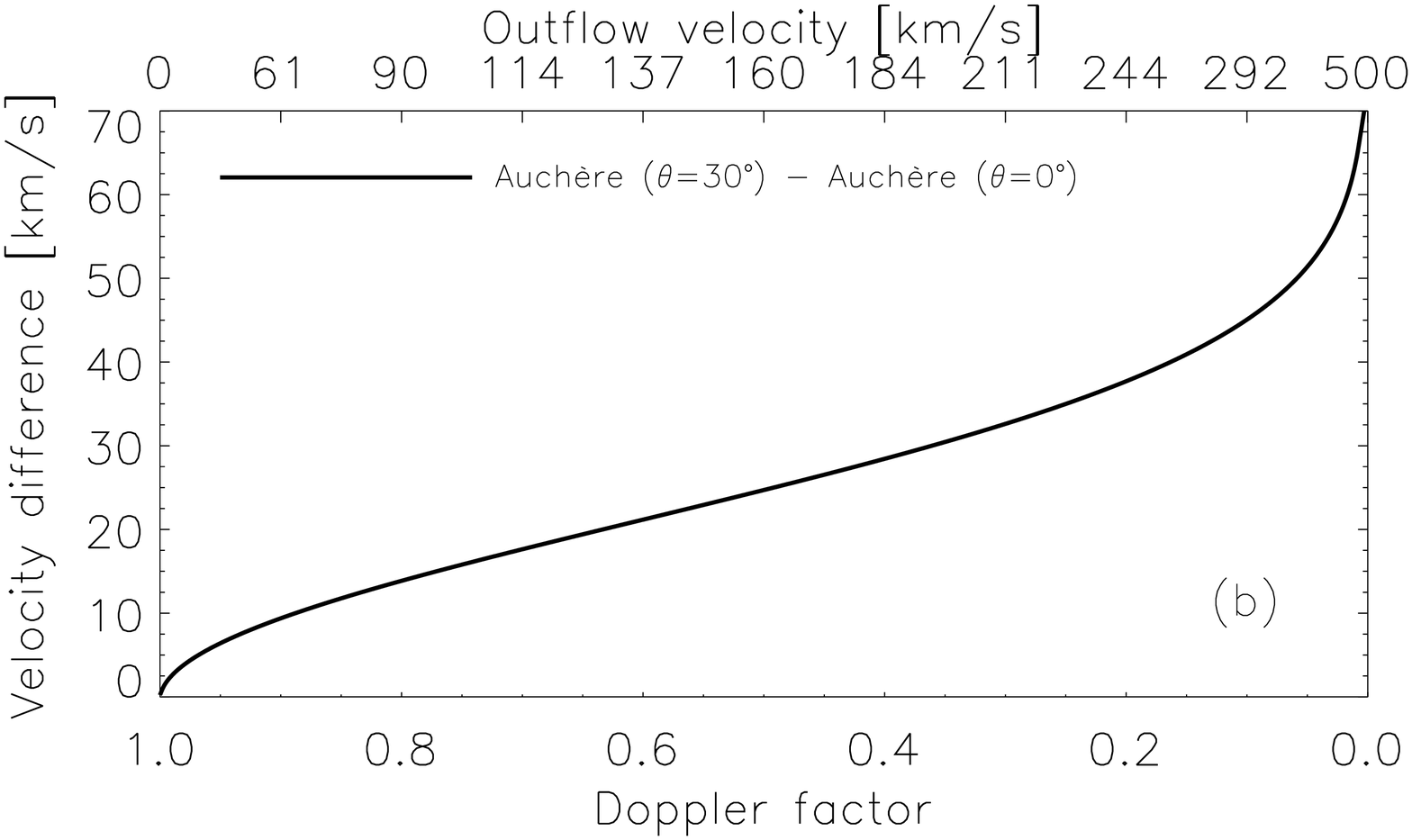}
        \caption{
        Panel ({\it a}): Doppler factor as a function of $v_w$ calculated considering the
        normalised \citet{Auchere2005} profile when $\theta = 0^\circ$ and $\theta = 30^\circ$.
        Triangles, squares, and asterisks: See Fig.~\ref{fig:DF}. Panel ({\it b}): Absolute
        values of the outflow velocity differences determined considering the normalised
        \citet{Auchere2005} profile when $\theta = 0^\circ$ and $\theta = 30^\circ$,
        as a function of the Doppler factor. The top axis scale shows the velocity values corresponding
        to the Doppler factor values when the \citet{Auchere2005} profile is taken into account and
        $\theta = 0^\circ$.
        }
        \label{fig:DF_theta_div}
\end{figure}

Furthermore, we performed a comparison considering different values of
the angle between the flow and the line-of-sight directions, $\theta$, using
the Auch\`ere profile. The results are shown in panel ({\it a}) of
Fig.~\ref{fig:DF_theta_div}, where we can note that assuming different angles
$\theta$ leads to the inference of remarkable variations of the \ion{H}{I} outflow velocity
values of about 14\%.

\begin{figure}[t]
        \centering
        \includegraphics[trim=0 82 0 0, clip, width=\hsize]{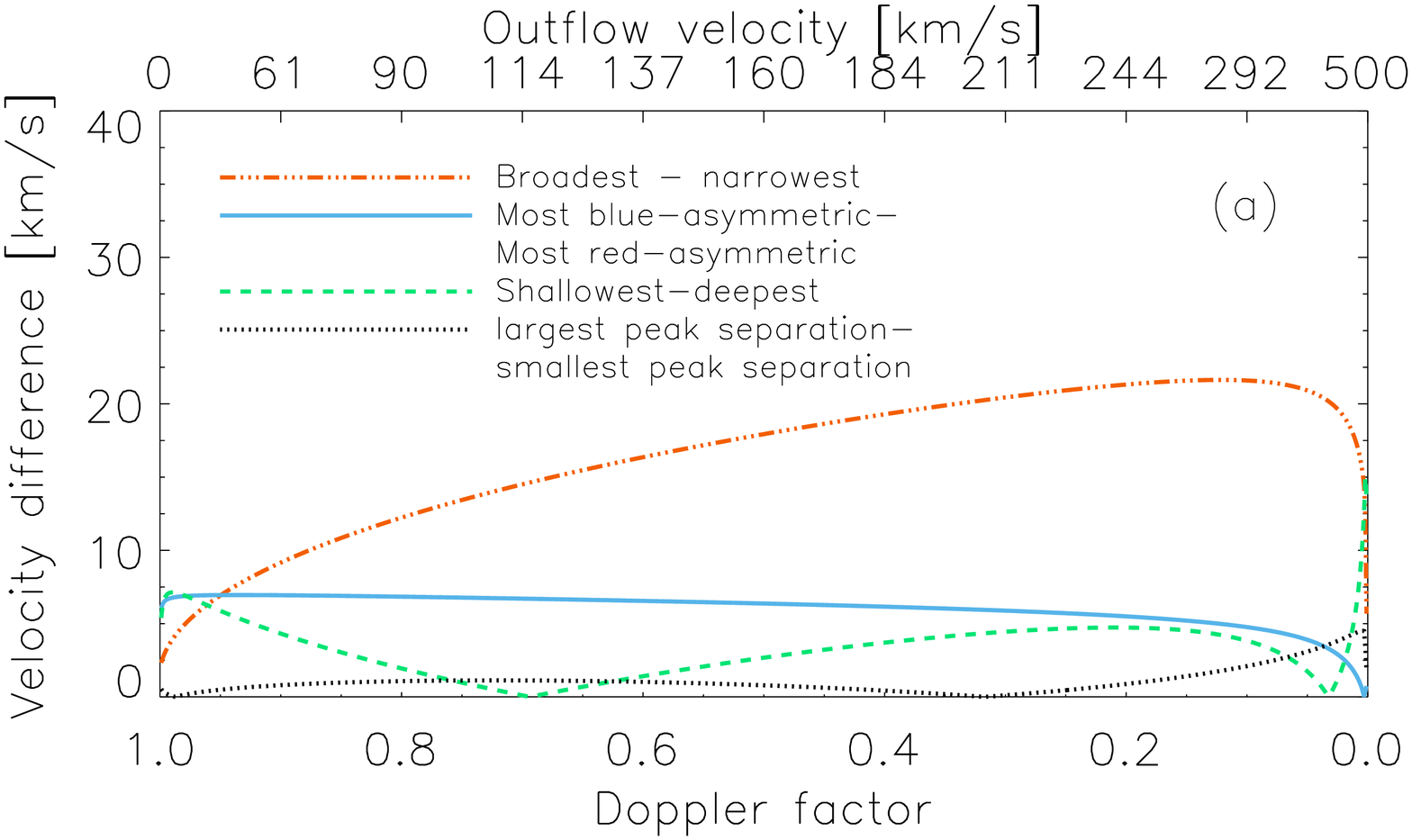}
        \includegraphics[trim=0 90 0 20, clip, width=\hsize]{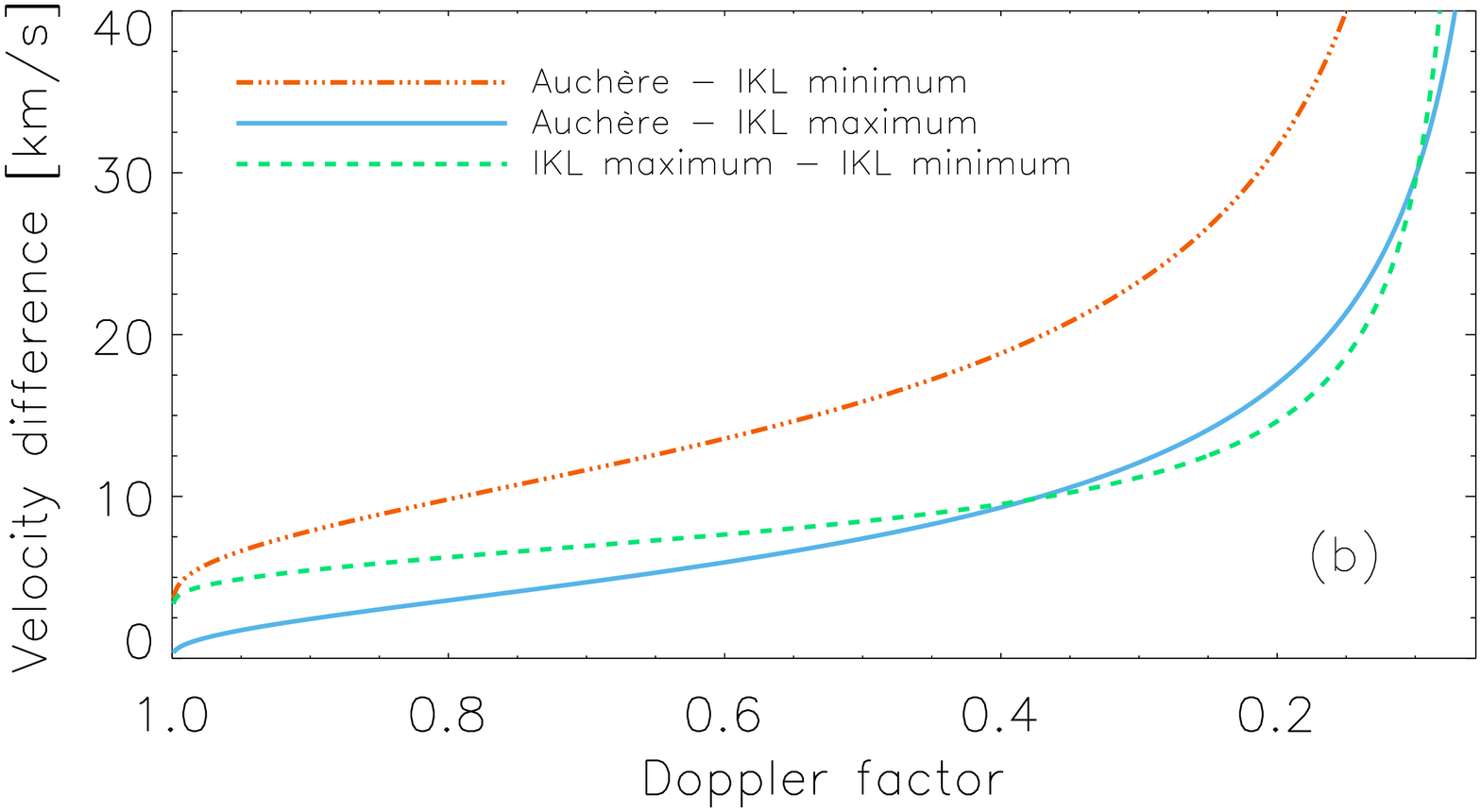}
        \includegraphics[trim=0 0 0 20, clip, width=\hsize]{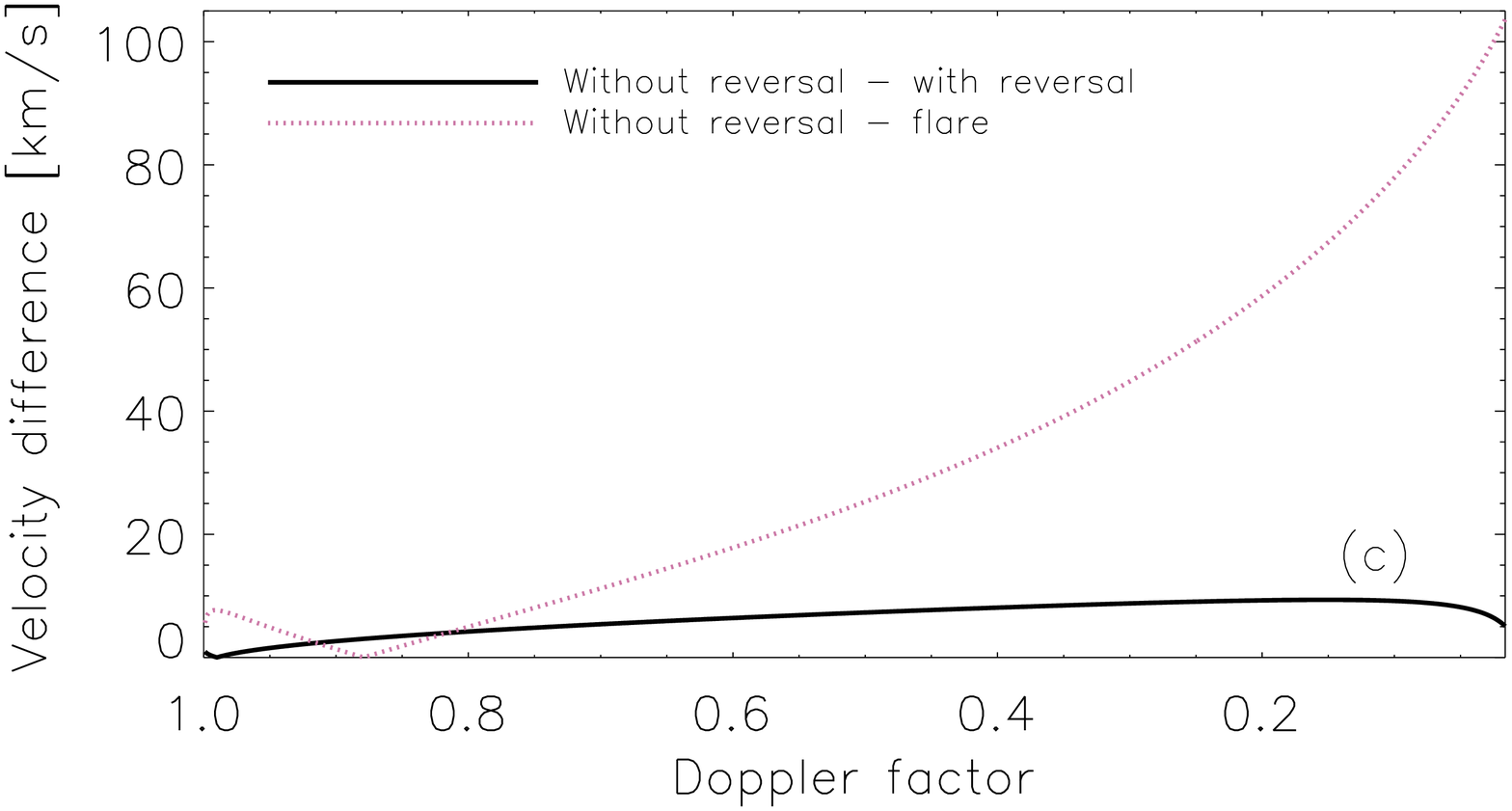}
        \caption{
        Panel ({\it a}): Absolute values of the differences between the outflow
        velocity values determined with the different observed chromospheric
        profiles shown in Fig.~\ref{fig:profili_osservati_param} as a function
        of the Doppler factor. The abscissa axis is complemented with the velocity
        values corresponding to the Doppler factor in the case of the \citet{Auchere2005}
        profile at the top of the panel. Panel ({\it b}): Same as in panel ({\it a}),
        but with the parametrised chromospheric profiles shown in
        Fig.~\ref{fig:profili_osservati_param} and in panel ({\it a}) of Fig.~\ref{fig:DF_prof_parametrizzati}.
        Panel ({\it c}): Same as in panel ({\it a}), but taking into account
        the profiles shown in panels ({\it a}) and ({\it c}) of
        Fig.~\ref{fig:ass_senz_ass_e_flare}.
        }
        \label{fig:velox_diffe_prof_osservati_e_param}
\end{figure}

To better understand these results, in Fig.~\ref{fig:velox_diffe_prof_osservati_e_param}
we plot the absolute value of the differences between the values of solar wind \ion{H}{I} velocity
corresponding to the same value of Doppler factor relative to the different chromospheric
observed and parametrised profiles. It is possible to note that in the former case
(observed profiles) the velocity differences are within a range of about
$22  \,\mathrm{km \,s}^{-1}$ when we take into account the extreme cases.
In the latter (parametrised profiles), the differences are within $30  \,\mathrm{km \,s}^{-1}$,
except for values of $D(v_w)$ close to zero, where the indetermination in the outflow
velocity is much higher. However, in both cases, up to values of $D(v_w) \approx 0.2$,
the relative velocity differences are below about 9\% and 12\%, respectively.
A similar result was obtained by \citet{Dolei2015}, who found that an uncertainty
within the range of \ion{H}{I} temperature measured by UVCS can return an error up to 10\% on
the determination of the theoretical coronal Ly$\alpha$ intensity, with an influence on
the estimate of the \ion{H}{I} outflow velocity up to $\pm 10-20  \,\mathrm{km \,s}^{-1}$.
When we consider the cases reported in Fig.~\ref{fig:ass_senz_ass_e_flare}, AR profiles
with and without reversal and a profile observed during a flare, we can see that for the
left panels of Fig.~\ref{fig:ass_senz_ass_e_flare} the velocity differences are below
$10  \,\mathrm{km \,s}^{-1}$, while for the right panels of Fig.~\ref{fig:ass_senz_ass_e_flare}
these values reach about $100  \,\mathrm{km \,s}^{-1}$ (see also Fig.~\ref{fig:velox_diffe_prof_osservati_e_param},
panel ({\it c})). Therefore, we find relative values of about 3\% and 21\%, respectively.
In Fig.~\ref{fig:DF_theta_div}, panel ({\it b}), concerning the Doppler factor curves relative
to different values of $\theta$, a similar result is shown. In this case, the absolute values
of the differences between the velocities corresponding to the same value of Doppler factor
are below $70  \,\mathrm{km \,s}^{-1}$, with relative values up to about 14\%.

Similarly to \citet{Dolei2016,Dolei2018,Dolei2019}, we created 2D maps of the solar wind
\ion{H}{I}
outflow velocity assuming the chromospheric Ly$\alpha$ line shape $\Psi(\lambda)$ as the sole
variable parameter,
while all the other inputs were kept fixed.
The maps have been created with a final
FOV ranging from $1.5 \,R_{\odot}$ to $3.95 \,R_{\odot}$.

\begin{figure*}[h]
        \centering
        \includegraphics[width=.5\hsize]{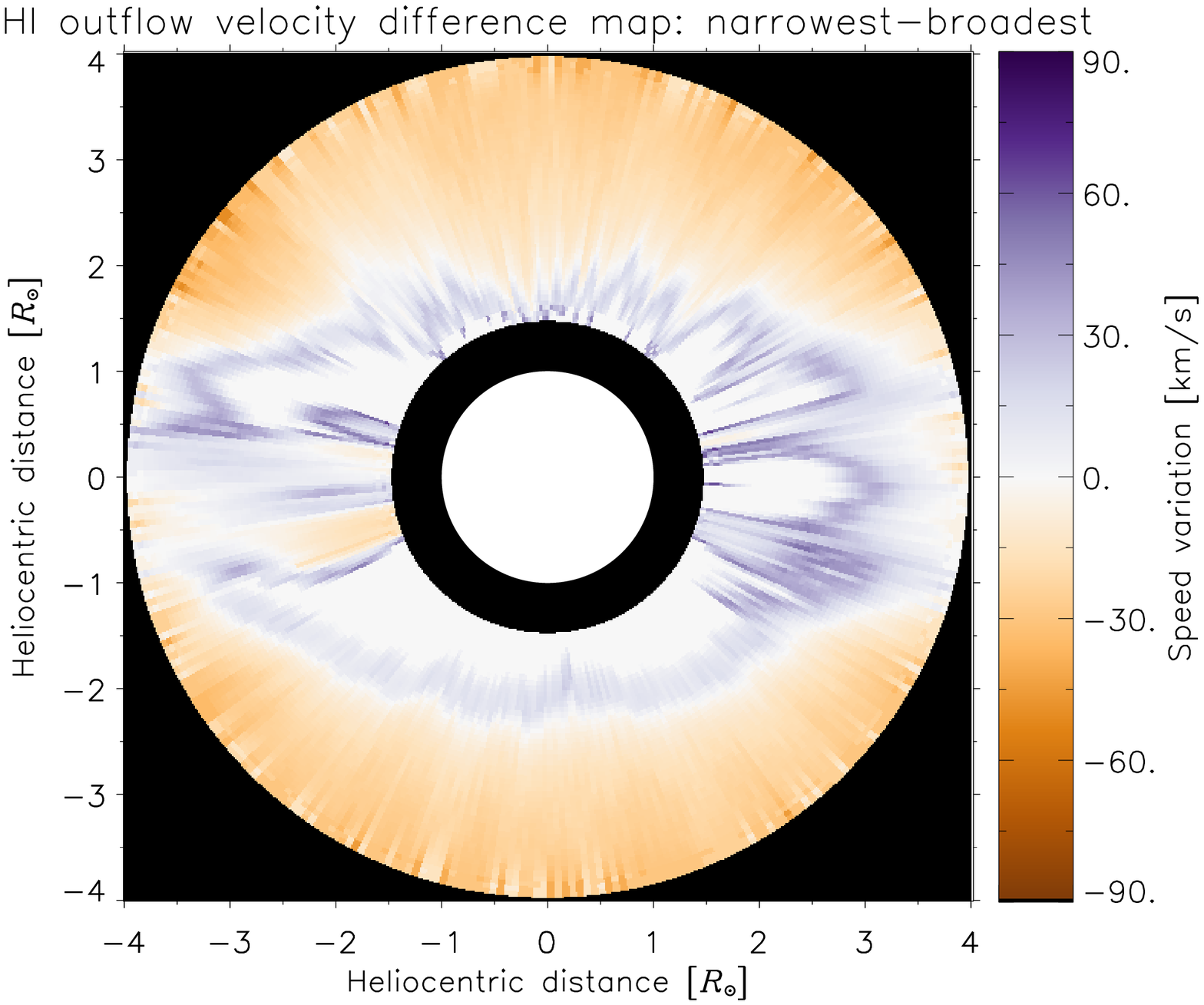}%
        \includegraphics[width=.5\hsize]{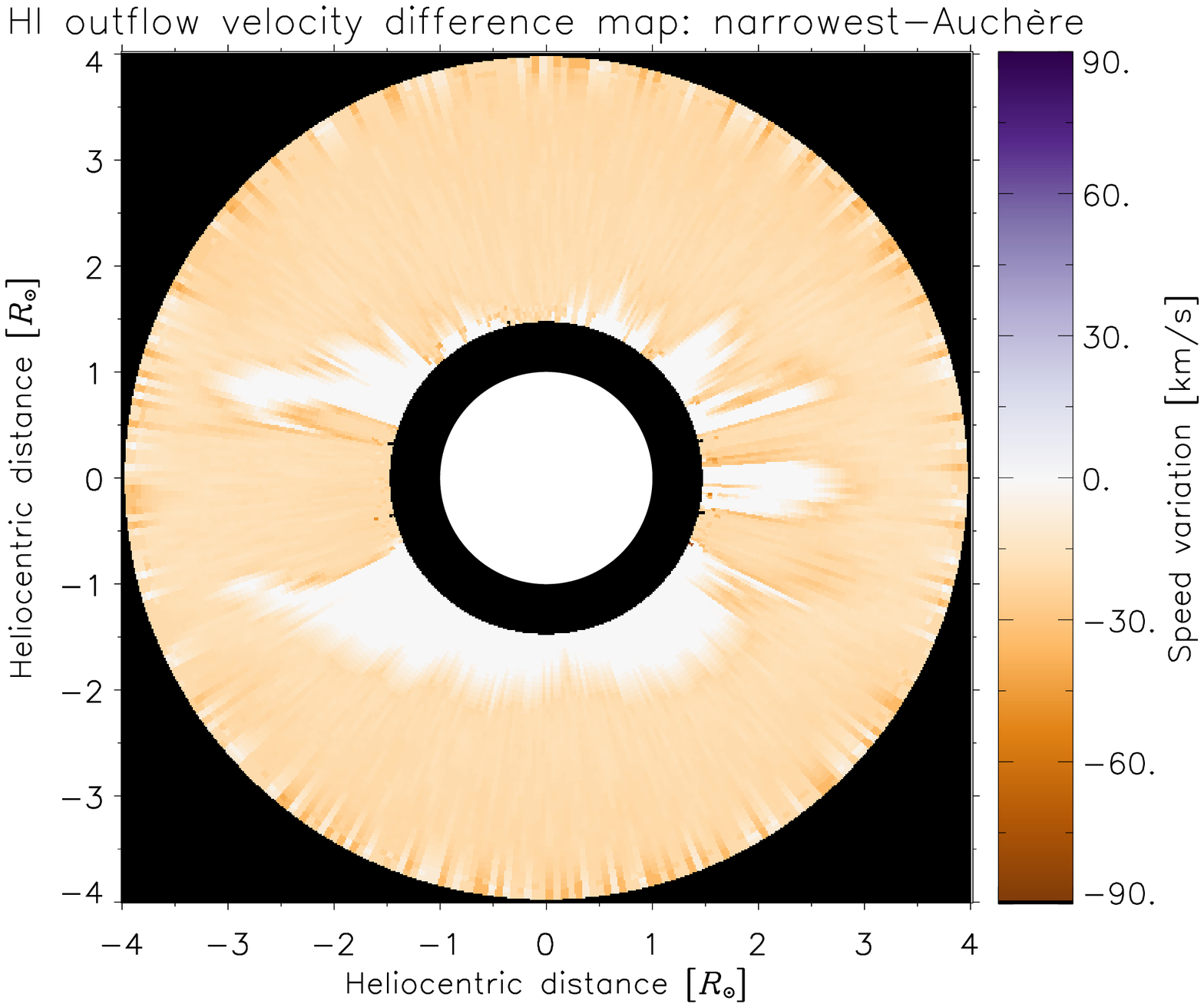}
        \includegraphics[width=.5\hsize]{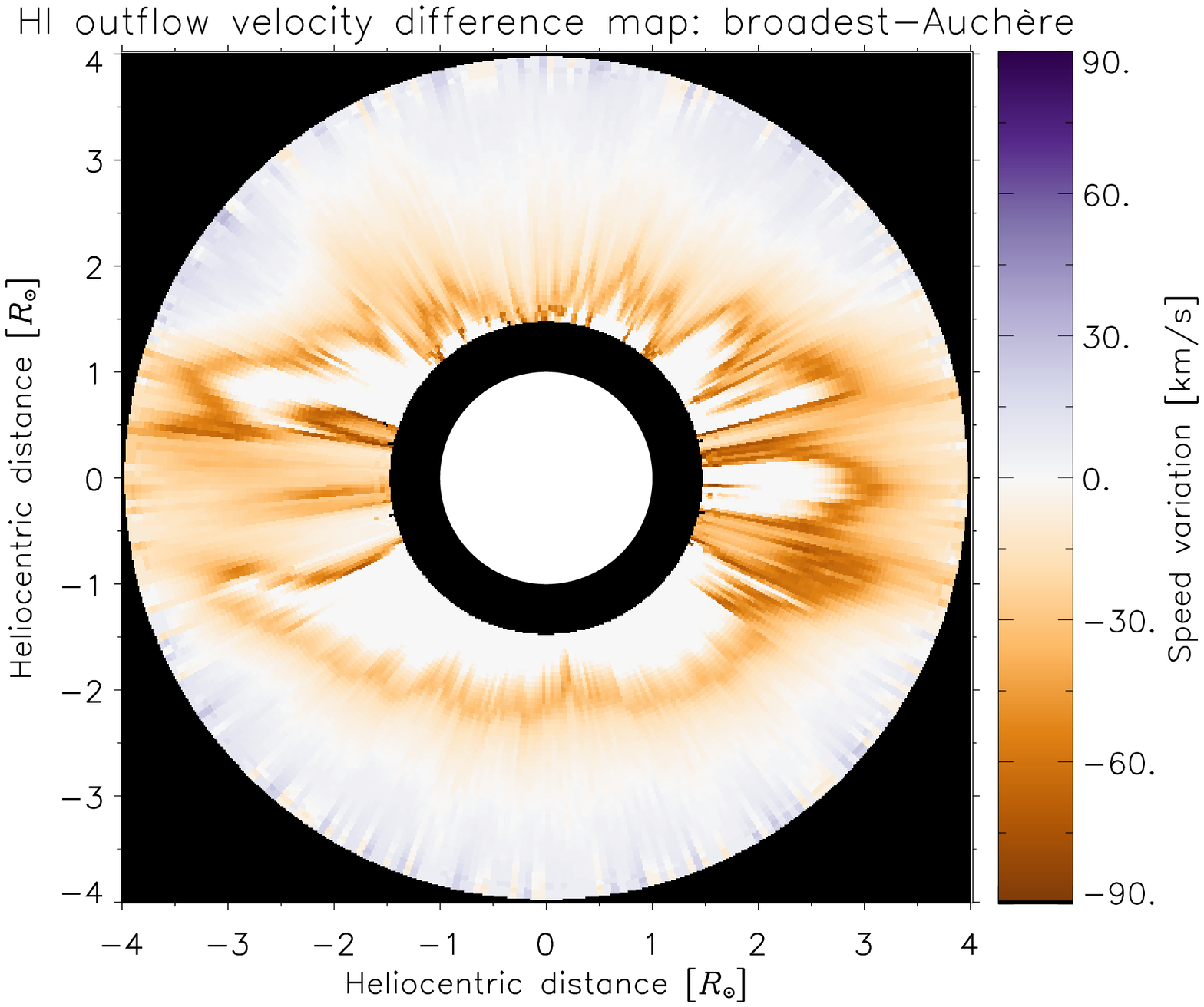}%
        \includegraphics[width=.5\hsize]{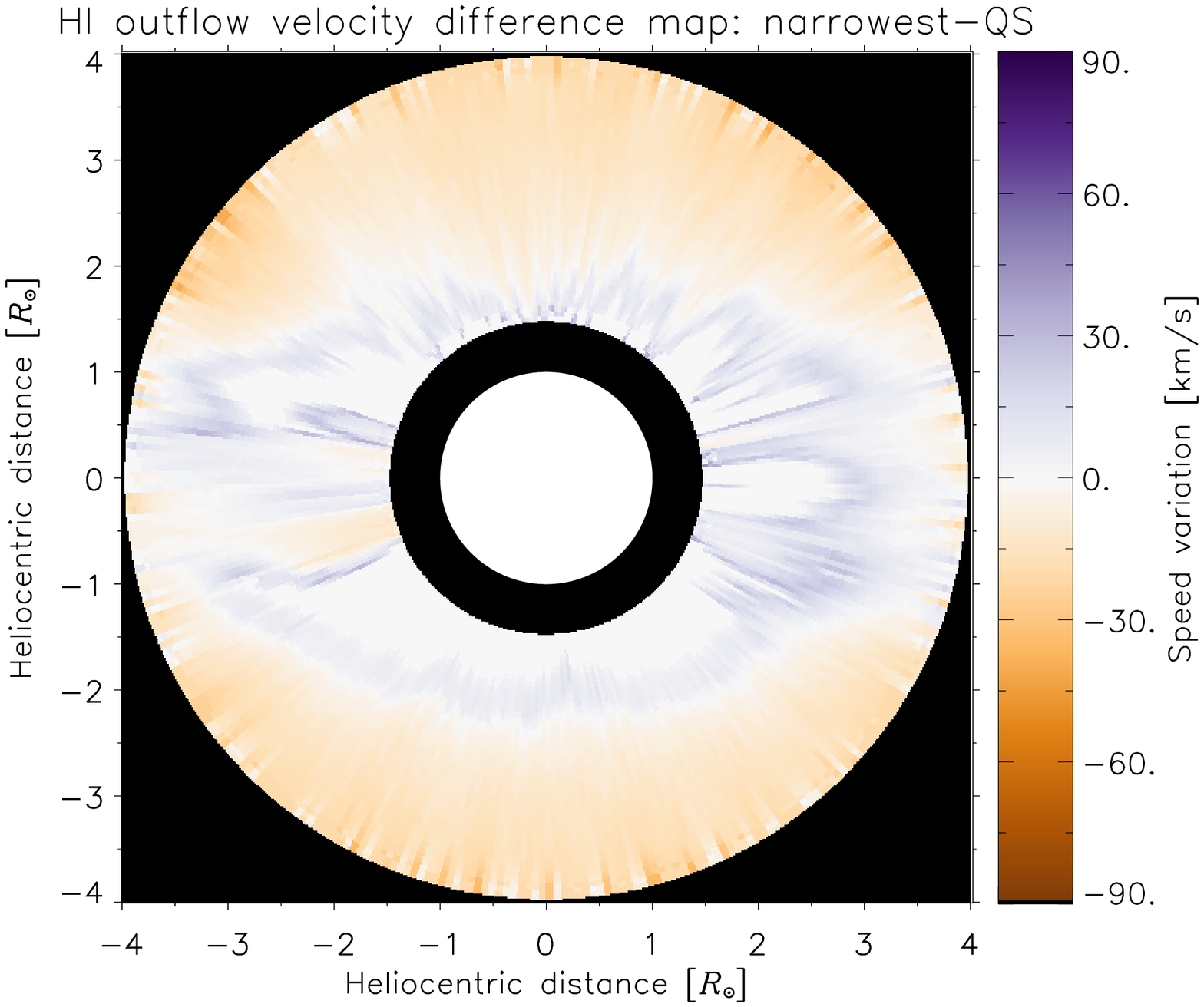}
        \caption{
        Differences between solar wind \ion{H}{I} outflow velocity maps obtained considering
        the narrowest profile \citep[AR 2340;][]{Fontenla1988} and the broadest one
        \citep[polar CH - April 17, 2009;][]{Tian2009b} (top left panel), the narrowest
        and the \citet{Auchere2005} analytic profiles (top right panel), the broadest and
        the \citet{Auchere2005} analytic profiles (bottom left panel), and the narrowest
        profile with that observed on September 23, 2008 \citep[QS;][]{Tian2009b} (bottom right panel).
        }
      \label{fig:diff}
\end{figure*}

For a better visualisation of the variations induced by the use of different
chromospheric line profile shapes, in Fig.~\ref{fig:diff}
we report the results obtained in terms of the differences between the outflow velocity
maps inferred from the various chromospheric profiles adopted: narrowest and
broadest profiles (AR 2340 and SUMER April 17, 2009 CH, respectively; top left panel),
narrowest and \citet{Auchere2005} profiles (top right panel), broadest and
\citet{Auchere2005} profiles (bottom left panel), and narrowest and QS
(SUMER, September 23, 2008) profiles (bottom right panel).
These difference maps exhibit RMS values equal to
$18  \,\mathrm{km \,s}^{-1}$, $19  \,\mathrm{km \,s}^{-1}$,
$20  \,\mathrm{km \,s}^{-1}$, and $11  \,\mathrm{km \,s}^{-1}$, respectively.
As can be seen in Fig.~\ref{fig:diff}, the differences are remarkable in the
left panels. In particular, in the bottom left panel the differences are larger
between about $1.5 \,R_{\odot}$ and $3.5 \,R_{\odot,}$ only in the equatorial region.

\section{Discussion and conclusions}
\label{sec:discus}

In this paper, we investigate the influence of the chromospheric
Ly$\alpha$ line profile on the determination of the solar wind \ion{H}{I}
outflow velocity through the Doppler dimming technique \citep*{HyderLites1970,Noci1987,Withbroe1982}.

In this analysis, we used spectroscopic measurements by SUMER
\citep{Lemaire2015,Curdt2008,Tian2009a,Tian2009b}, which include the data
sets with the highest resolution of Ly$\alpha$ solar-disc observations
presently available. We also selected some profiles of ARs and QS acquired
by UVSP and described in \citet{Fontenla1988}, where the correction concerning
the geocoronal absorption was applied, along with a profile reported in
\citet{Lemaire1984} characterising a solar flare, and another one acquired
in an equatorial CH \citep{BocchialiniVial1996}; both were obtained by the LPSP
instrument on OSO-8. For the sake of completeness, it is worth recalling
that the Chromospheric Lyman Alpha Spectropolarimeter (CLASP) observed the
Sun in \ion{H}{I} Ly$\alpha$ during a suborbital rocket flight in 2015
\citep{CLASP}. These observations measured the Ly$\alpha$ profile in
a quiet-Sun target near the limb with a slit 400\arcsec{} long and 1\farcs44 wide.
Overall, these observed profiles were in agreement with the early measurements of
\citet{Gouttebroze1978}, but they were affected by geocoronal absorption and water vapor
contamination. Furthermore, SUMER measurements were recently used to derive
a reference QS Ly$\alpha$ profile that would be representative of the
Ly$\alpha$ radiation from the solar disc during a minimum solar activity \citep{Gunar2020}.

We analysed the behaviour of the Doppler factor with respect to the
variation of four parameters on which the pumping Ly$\alpha$ profile depends:
line width, reversal depth, asymmetry of the peaks, and separation of the peaks.
Moreover, taking into account the same input data adopted by \citet{Dolei2019},
that is, $T_e$, $T_{\ion{H}{I}}$, $n_e$, and the Ly$\alpha$ coronal
intensity, and using a constant value for the radiance of the exciting chromospheric
Ly$\alpha$ radiation, we computed outflow velocity maps, where the sole
variable is the profile of the exciting chromospheric Ly$\alpha$ radiation.

It is worth mentioning that just recently
\citet{Cranmer2020} obtained radial profiles of electron and proton
temperatures very different to those adopted in the present work. Since
our main goal here is to investigate the effects of the chromospheric
Ly$\alpha$ profile shape on the determination of the outflow velocity,
we feel confident that the evaluation of Cranmer's temperature profiles
can be deferred to a future analysis without questioning the results
obtained in the present work.

Considering chromospheric Ly$\alpha$ normalised profiles observed by
SUMER, UVSP, and the LPSP instrument on OSO-8, we measured their
FWHM, reversal depth, asymmetry of the peaks, and separation of the peaks.
We found a maximum variation of about 50\%, 69\%, 35\%, and 50\%
between the maximum and minimum values of each parameter on which the
Ly$\alpha$ profile depends, respectively, referring each percentage
to the minimum value of the corresponding parameter.
Using these observations, we found that the Doppler factor $D(v_w)$,
which is a measure of the overlapping between the coronal and chromospheric
profiles as a function of the \ion{H}{I} outflow velocity, does not strongly depend on the observed parameters that characterise the chromospheric
Ly$\alpha$ profiles. In fact, except the case in which we consider a flare,
the relative velocity differences determined using the different observed
profiles, as a function of the Doppler factor, are below about 9\%, a value
that is comparable with the uncertainty in the velocity values determined
by the Doppler dimming technique. 

In order to further illustrate this effect, we also considered other
analytic profiles, which were proposed by \citet{Auchere2005} and \citet{ikl2018}.
Also in these cases, we note negligible effects on the Doppler
factor evaluation, obtaining relative absolute values of the velocity
differences within a range of about 12\% in correspondence of the same
values of the Doppler factor. Conversely, non-negligible differences appear
in the Doppler factor when we take into account a flare profile.
In fact, in this case the resulting velocity variation is significantly larger,
increasing up to about $100  \,\mathrm{km \,s}^{-1}$ and returning a
relative velocity difference values of about 21\%
(see Fig.~\ref{fig:velox_diffe_prof_osservati_e_param}, panel ({\it c})).

We also show the effect of considering different values of the angle
between the flow and the line-of-sight directions, $\theta$. We find that
if $\theta > 0^\circ$, the Doppler factor curve has a smoother trend than
for $\theta=0^\circ$. Figure~\ref{fig:DF_theta_div} shows the case of
$\theta = 30^\circ$. Therefore, when we consider a fixed value of the Doppler factor,
the corresponding velocity value is greater when $\theta > 0^\circ$.
However, the velocity differences remain below about $70  \,\mathrm{km \,s}^{-1}$
in this case, with a relative value equal to $\approx 14\%$. It is worth
mentioning that such a value for $\theta$ can be obtained only for very
close distances from the Sun, of about $2 R_{\odot}$, so its
contribution in the computation of the \ion{H}{I} outflow velocity is
limited; this remains more valid for higher values of $\theta$.

We used the narrowest and the broadest profiles, the one observed on
September 23, 2008 (QS), and the analytic profile deduced by \citet{Auchere2005},
to derive the 2D maps of \ion{H}{I} outflow velocity. Taking into account the
velocity difference maps shown in Fig.~\ref{fig:diff}, we found that RMS
values of the difference of the solar wind \ion{H}{I} speed are equal to
$18  \,\mathrm{km \,s}^{-1}$ (narrowest - broadest), $19  \,\mathrm{km \,s}^{-1}$
(narrowest - Auch\`ere), $20  \,\mathrm{km \,s}^{-1}$ (broadest - Auch\`ere),
and $11  \,\mathrm{km \,s}^{-1}$ (narrowest - QS). These values can be
considered as possible uncertainties in the estimate of the outflow velocity
with regard to the dependence on the chromospheric profile shape.
However, they are significantly smaller than those found by \citet{Dolei2018},
which are related to other parameters. In fact, assuming a maximum uncertainty on the other
physical quantities of $\pm 30$\%, they estimated that the resulting uncertainties
on the derived velocity were of $67  \,\mathrm{km \,s}^{-1}$,
$67  \,\mathrm{km \,s}^{-1}$, $81  \,\mathrm{km \,s}^{-1}$,
and $45  \,\mathrm{km \,s}^{-1}$ for the impact of electron density,
total chromospheric intensity, electron temperature, and \ion{H}{I} temperature, respectively. 

Therefore, our results indicate that even the largest variations actually observed
in the parameters on which the chromospheric Ly$\alpha$ profile depends,
related to the solar magnetic activity and to different disc regions,
return small differences in the \ion{H}{I} outflow velocity estimate,
excluding the flare case, where non-negligible effects are present.
This reveals how little effect the shape of the exciting chromospheric Ly$\alpha$
profile has on the determination of the solar wind \ion{H}{I} outflow velocity with
respect to the other parameters characterising the scattered coronal line.
As a consequence, a unique shape of the Ly$\alpha$ chromospheric profile can
be adopted all over the solar disc; moreover, analytical chromospheric profiles
can be used, such as those proposed by \citet{Auchere2005} and \citet{ikl2018},
without significantly affecting the solar wind \ion{H}{I} velocity computation
beyond the uncertainties characterising the Doppler dimming technique.

Data coming from the Metis coronagraph \citep{Antonucci2020, Fineschi2020},
aboard the Solar Orbiter spacecraft \citep{Muller2020}, will return more
detailed and accurate information able to update this and previous studies,
thanks to simultaneous UV and polarised VL observations with high spatial
and temporal resolution.

\begin{acknowledgements}
The authors acknowledge the support of the Italian Space Agency (ASI) to this
work through contracts ASI/INAF No. I/013/12/0 and No. 2018-30-HH.0.
The authors thanks the anonymous referee for useful comments and suggestions,
which led to a sounder version of the manuscript.

\end{acknowledgements}

\end{document}